\documentclass[sigconf,nonacm,screen,balance=false]{acmart}
\setcopyright{none}
\acmConference{}{}{}
\acmDOI{}
\acmISBN{}
\acmBooktitle{}
\date{}
\AtBeginDocument{}

\begin{document}
\title[Your Programming Students\textquotesingle{} Cognition with ChatGPT]{Your Programming Students\textquotesingle{} Cognition with ChatGPT: Higher Performance, Lower Retention, and Reduced Ownership}

\author{Christian Bergh}
\email{c.bergh@unsw.edu.au}
\affiliation{%
  \institution{University of New South Wales}
  \city{Sydney}
  \state{NSW}
  \country{Australia}
}
\author{Benjamin Tag}
\email{benjamin.tag@unsw.edu.au}
\affiliation{%
  \institution{University of New South Wales}
  \city{Sydney}
  \state{NSW}
  \country{Australia}
}
\author{Alexandra Vassar}
\email{a.vassar@unsw.edu.au}
\affiliation{%
  \institution{University of New South Wales}
  \city{Sydney}
  \state{NSW}
  \country{Australia}
}
\author{Jake Renzella}
\email{jake.renzella@unsw.edu.au}
\affiliation{%
  \institution{University of New South Wales}
  \city{Sydney}
  \state{NSW}
  \country{Australia}
}

\renewcommand{\shortauthors}{Bergh et al.}

\begin{abstract}
Generative AI can improve students' programming performance, but successful task completion may not reflect what they retain. We examined performance, retention, cognitive load, and ownership in a controlled between-subjects experiment with 59 undergraduate computer science students, 55 were retained for analysis. Participants completed three introductory C programming tasks with access to ChatGPT-4.5 or conventional web search without generative AI. We measured task performance, self-reported mental effort and difficulty, pupillary responses, heart rate variability, and ownership, and assessed cued recall immediately and 48 hours later. ChatGPT-assisted students achieved higher coding scores (89\% vs. 69\%) but lower recall scores immediately (41\% vs. 53\%) and after 48 hours (39\% vs. 52\%). There was no significant difference in the loss of recall information over 48 hours between the groups. Self-reported mental effort increased less across tasks in the ChatGPT condition (Holm-adjusted p = .047), and students attributed less of the submitted code to themselves (45\% vs. 81\%). Confirmatory physiological tests did not detect significant differences in trajectories between conditions; substantial data loss limits their interpretation. These findings reveal a gap between assisted task performance and subsequent recall and sense of ownership in this setting. They motivate the need for assessment practices and AI learning tools that require students to explain, retrieve, and contribute to the work they submit as active participants in their education.
\end{abstract}

\keywords{Generative AI, ChatGPT, Computing Education, Cognitive Load, Knowledge Retention, Psychological Ownership, AI-Assisted Programming}

\maketitle

\section{Introduction}
Since the release of ChatGPT in November 2022, generative Artificial Intelligence (GenAI) has become part of students' routine; they use large language models (LLMs) to draft essays, debug code, and complete assignments~\cite{Naznin2025ChatGPTReview}, and many tertiary institutions have replaced initial prohibition with policies permitting disclosed use~\cite{Qian2026GoverningUniversities,Parker2025ComparativeCountries,Geng2026MappingAnalysis}. 
These changes are consistent with calls from UNESCO and others to prepare students for an AI future~\cite{Pedro2019ArtificialDevelopment}. Educators have traditionally relied, in part, on students demonstrating knowledge acquisition through summative assessment tasks. This practice assumes that the artefact is a sufficient proxy to measure a student's knowledge or capability. LLMs can produce a plausible essay or functional code without the student acquiring that knowledge. Research to date has established that GenAI assistance improves immediate task performance~\cite{Wu2025Human-generativeMotivation, Noy2023ExperimentalIntelligence}, but there is limited evidence on what students actually retain from AI-assisted work~\cite{Kosmyna2025YourTask,Lee2025TheWorkers}.  

Much of the current research focuses on technical aspects such as model performance~\cite{Bommasani2021OnModels, Liang2022HolisticModels}, methods for integrating GenAI in specific use cases~\cite{Mustafa2024AAgenda}, and detection of AI-generated content~\cite{Wu2025ADirections}. Comparatively little work examines the cognitive impacts of GenAI on users, especially in a higher education setting~\cite{Deng2025DoesStudies}. AI-assisted learning may also reshape learners' preconceptions of their capabilities and ownership of the work they produce.

We conducted a between-subjects controlled experiment exploring the cognitive impact of GenAI usage on $N=59$ undergraduate computer science students, of whom $n=55$ were retained for analysis. These students were tasked with completing introductory-level coding problems with or without GenAI assistance. During the programming tasks, we recorded pupillometry and heart rate as physiological metrics of cognitive load, alongside self-reported measures. After completing the problems, students reported their sense of ownership over the code they had submitted. Retention was assessed by quizzing students on the content of the coding problems immediately after submission and again 48-hours later. The AI-assisted group scored significantly higher on the coding assessment ($89\%$ vs. $69\%$). However, they scored significantly lower on both retention quizzes: immediately ($41\%$ vs. $53\%$) and 48~hours later ($39\%$ vs. $52\%$). In addition, they reported a smaller increase in mental effort across tasks ($p_\text{Holm}=.047$) and a significantly lower sense of ownership over the submitted code ($45\%$ vs. $81\%$). This experiment allowed us to evaluate the immediate and delayed effects of GenAI assistance on learners' performance, cognitive load, retention, and sense of ownership over their work. 

Understanding these effects is increasingly important as GenAI tools become embedded in learning environments. Investigating these interactions is critical for determining the role GenAI plays in educational processes and its impact on learning outcomes. Specifically, how can programming courses assess learning in a post-AI world?

Toward this goal, this paper contributes: 
\begin{itemize}
    \item A combined physiological and self-report account of cognitive load during programming tasks with or without GenAI assistance.
    \item Evidence of a dissociation between short-term task success and longer-term (48 hours later) knowledge retention in AI-assisted learning.
    \item Evidence that GenAI assistance lowers students' sense of ownership over submitted work. 
    \item Design implications and suggested mitigations for AI-assisted learning systems that aim to balance performance support with durable learning at the institution, educator, and student levels.
\end{itemize}

\section{Related Work}
Generative AI (GenAI) is increasingly used in education to support translation, personalisation, feedback, and content generation, while concerns about inequitable access remain~\cite{Roscoe2022InclusionEducation, Ghosh2024ChatGPTClasses, Holstein2021EquityEducation, Srinivasa2022HarnessingEducation}. It has also raised concerns about academic integrity because students can produce substantial amounts of text or code with minimal input, allowing them to submit work that may not reflect their own learning or understanding~\cite{Nguyen2024UnmaskingUndergraduates, Oravec2023ArtificialBard}. While these concerns are often framed as issues of misconduct, they may also reflect a broader educational trade-off in which GenAI assistance reduces cognitive effort while potentially affecting knowledge retention and durable learning. To situate this work, we review the literature on GenAI in computing education, cognitive load and learning, the performance and cognitive impacts of AI-assisted learning, and questions of ownership and authorship in human-AI creation.

\subsection{GenAI in Computing Education}
Early work highlighted the disruptive potential of GenAI for programming education, as models rapidly approached or exceeded novice programming capabilities~\cite{Finnie-Ansley2022TheProgramming, Finnie-Ansley2023MyExercises}. As the initial capabilities of GenAI code generation systems matured, the discussion shifted to investigating the extent to which students should be allowed to use GenAI and how to use it effectively. \citet{Denny2024ComputingAI} argue that experts derive value from GenAI precisely because they possess the foundational knowledge to analyse, critique, and contextualise generated code. The authors further argue for an evolution of computing education pedagogical and assessment practices to reflect an increasingly AI-assisted computing world~\cite{Denny2024ComputingAI}.

This evolving discussion of disruption, adaptation, and pedagogy establishes the computing education context for empirical investigations into the impact of GenAI assistance on novice programming learners. One of the earliest studies examined the effect of GenAI assistance on novice programmers using a web-based application for learning introductory Python~\cite {Kazemitabaar2023StudyingProgramming}. The authors found that GenAI significantly improved coding performance but did not significantly affect retention evaluation either immediately or one week later. However, while this study demonstrated performance benefits, it did not investigate how GenAI assistance may have impacted underlying cognitive processes. Additional studies in the field have begun investigating the effects of GenAI assistance on cognitive processes. 

\subsection{Learning and Cognitive Load}
Learning science research shows that activities requiring learners to generate and retrieve information often produce stronger long-term retention than conditions that maximise immediate performance~\cite{Soderstrom2015LearningReview,Slamecka1978ThePhenomenon.,Roediger2011TheRetention,Bjork2020DesirablePractice}. This creates a potential tension for AI-assisted learning, as GenAI can improve task performance while reducing opportunities for productive cognitive effort~\cite{Bjork2011MakingLearning}. 

Cognitive Load Theory provides a complementary account; it posits that learning requires time and conscious effort through a limited-capacity working memory before it can be stored in long-term memory~\cite{Sweller2019CognitiveTechnology}. Load on working memory is conventionally decomposed into three categories: intrinsic load, imposed by the learning material itself; extraneous load, imposed by how the material is presented; and germane load, characterised as the effortful processing through which schema construction occurs~\cite{Sweller1998CognitiveDesign}. Suggesting that learning benefits arise not simply from reducing extrinsic cognitive load, but from preserving the germane processing required for schema construction and long-term retention~\cite{Sweller2019CognitiveTechnology, Sweller1998CognitiveDesign, Sweller2011CognitiveTheory}.

\subsection{Performance and Cognitive Impacts of AI-Assisted Learning}
The evidence that GenAI assistance improves immediate outcomes is consistent. In introductory programming, students given access to ChatGPT reported higher programming self-efficacy and motivation and scored higher on computational thinking than a control group~\cite{Yilmaz2023TheMotivation}. \citet{Wang2025ImpactCourses} report similar findings, students collaborating with a GenAI agent outperformed a conventional computer-supported collaborative learning group in achievement, self-efficacy, and interest, and reported significantly lower mental effort with no difference in mental load. A meta-analysis of GenAI tools in programming education pooled these effects across studies, AI-assisted students completed tasks faster and achieved higher task performance scores, and perceptions of the tools were almost uniformly positive~\cite{Alanazi2025TheMeta-Analysis}. Notably, the same analysis found no significant advantage in learning success or ease of understanding~\cite{Alanazi2025TheMeta-Analysis}. These benefits focus on task output, speed, and how learners feel about their work, and do not cleanly extend to what they learn and retain over time.

This divergence between performance and learning becomes particularly visible when researchers examine what students retain, understand, or can independently reason about; rather than the quality of their immediate outputs. Across domains, GenAI support consistently reduces cognitive effort and improves subjective experiences while producing more mixed outcomes for learning~\cite{Stadler2024CognitiveInquiry, Li2025GenerativeProcesses, Prather2023TheEducation}. Students using GenAI report lower cognitive load, greater confidence, and stronger perceptions of success, yet often demonstrate weaker reasoning, poorer argumentation, and reduced knowledge transfer when assessed beyond the immediate task \cite{Stadler2024CognitiveInquiry, Li2025GenerativeProcesses}. Programming studies suggest that these benefits are not distributed evenly. More capable learners tend to use GenAI as a collaborative resource, critiquing suggestions and integrating them into their own reasoning, whereas struggling learners are more likely to accept outputs uncritically and become dependent on the tool \cite{Li2025GenerativeProcesses,Prather2024TheProgrammers}. As a result, GenAI can simultaneously improve task completion while limiting the cognitive engagement necessary for durable learning.

Evidence from educational and workplace settings points to a common underlying mechanism: cognitive offloading. Learners who rely heavily on GenAI may invest less effort in generating, evaluating, and refining solutions themselves, instead shifting their attention toward monitoring and integrating AI-generated content \cite{Lee2025TheWorkers}. This reduction in active processing has been linked to weaker transfer, overconfidence, and illusions of competence, where users believe they understand more than they actually do \cite{Prather2024TheProgrammers}. Physiological evidence further supports this interpretation. Using EEG recordings collected over four months, \citet{Kosmyna2025YourTask} found that participants writing with GenAI exhibited the lowest levels of neural engagement, reported the weakest sense of ownership over their work, and struggled to recall or quote content they had produced. Together, these findings suggest that GenAI assistance does not merely reduce workload; it can also alter the extent to which learners remain cognitively and psychologically invested in the task. 

\citet{Liu2026ToolEducation} conceptualise these outcomes as a tension between \emph{Domain Mastery} and \emph{Tool Mastery}. Depending on factors such as task complexity, time pressure, and scaffolding design, learners may enter either a \emph{Scaffolding Loop}, in which GenAI supports and extends their own reasoning, or an \emph{Offloading Loop}, in which routine delegation displaces opportunities for germane processing, knowledge construction, and authorship \cite{Liu2026ToolEducation}. While prior work collectively links GenAI use to lower cognitive effort, reduced critical engagement, weaker transfer, and diminished ownership \cite{Kazemitabaar2023StudyingProgramming,Wang2025ImpactCourses,Li2025GenerativeProcesses,Prather2024TheProgrammers,Yilmaz2023TheMotivation}, no study to our knowledge has simultaneously examined performance, cognitive load, retention, and psychological ownership in novice programmers using AI-assisted programming.

\subsection{Ownership and Authorship in Human--AI Creation}
Psychological ownership describes a relation between an individual and an object, material or immaterial, in which the object comes to be experienced as an extension of the self~\cite{Olckers2017MeasuringReview}. Psychological ownership has been linked to individual attitudes and behaviour, and validated scales exist for measuring it across domains~\cite{Olckers2017MeasuringReview}. 

Research on ownership of AI-generated content remains limited. Wasi et al.~\cite{Wasi2024LLMsReasoning} found that ownership varies based on content type, with individuals feeling less ownership over creative content, such as poems and more over technical content, such as emails~\cite{Wasi2024LLMsReasoning}. The same work also identified a possible disconnect between ownership and authorship, with individuals willing to claim authorship of content they did not feel they owned~\cite{Wasi2024LLMsReasoning}. Draxler et al. report a related reluctance to attribute authorship to AI-generated content~\cite{Draxler2024TheAuthors}. Research involving ownership of AI-generated content should therefore capture individuals' claims of authorship alongside ownership, since the two seem to exist independently. 

\subsection{Research Question}
Existing work shows that GenAI assistance can improve programming performance, but performance gains do not necessarily translate into learning. Prior studies have reported lower cognitive effort, poorer transfer, reduced critical engagement, and differences in these effects across learners~\cite{Kazemitabaar2023StudyingProgramming, Wang2025ImpactCourses, Li2025GenerativeProcesses, Prather2024TheProgrammers}. Related work has also linked GenAI assistance to reduced recall and ownership in writing tasks~\cite{Kosmyna2025YourTask}. However, no work to our knowledge jointly examines performance, cognitive load, retention, and psychological ownership in AI-assisted programming.

In this paper, we address these gaps through the following research questions. 

\begin{enumerate}
    \item [\textbf{RQ1}] How does the use of GenAI tools shape learners' performance, cognitive load, retention, and psychological sense of ownership in introductory programming tasks? 
    \begin{enumerate}
        \item How does GenAI tool usage affect task performance in introductory programming assignments, as measured by accuracy and completeness?
        \item How does GenAI tool usage affect learners' cognitive load during programming tasks, as measured through subjective and physiological indicators?
        \item How does GenAI tool usage influence immediate and delayed retention of information in the programming tasks?
        \item How does GenAI tool usage affect learners' sense of ownership over the code they produce?
    \end{enumerate}
\end{enumerate}

\section{Method}
We conducted a between-subjects, mixed-methods experiment examining the cognitive, psychological, and physiological effects of GenAI use during a programming task. Participants completed an introductory programming assessment either with access to a GenAI assistant or without. The task was designed to mimic a student completing a programming homework assessment, since this is a common setting in which students interact with GenAI tools while producing assessable work. 

\subsection{AI Disclosure}
\label{sec:ai-disclosure}
The authors used GenAI tools to improve language and readability in the manuscript (Microsoft 365 Copilot). The statistic pipeline was developed using Claude Code via Fable 5.1. All outputs were reviewed and verified by the authors (Authors tested all generated code, reviewed code outputs for accuracy including validating results with external analysis (SPSS v30.0.0), and reviewed any edits for accuracy). GenAI (ChatGPT Astra) was used to anonymise experimental setup images~\autoref{fig:teaser}. Authorship boundaries: GenAI did not make substantive intellectual contributions; authors retain full responsibility for ideas, interpretations, and claims. No references were identified or generated using GenAI tools.

\subsection{Human Research Ethics}
This study was conducted under the approval of our university's Institutional Review Board (IRB), ensuring adherence to ethical research standards. The study procedure was reviewed and approved by the IRB. All participants provided written informed consent before taking part, were informed that they could withdraw at any point without penalty, and were provided with a reimbursement gift voucher for their time. Physiological data were recorded only for the duration of the experimental session. All data were stored in university-managed secure storage and analysed in de-identified form.

\subsection{Participants}
We recruited $N=59$ undergraduate Computer Science students at a single university. The students were pseudo-randomly assigned to the experimental AI-Assisted ($n=31$) and No AI groups ($n=28$) in order to ensure gender balance across conditions. The AI-Assisted group comprised $7$ females and $24$ males; the No AI group comprised $6$ females and $22$ males. No participant identified as a third gender or declined to provide their gender. 

In order to balance programming experience, participants were recruited from first year undergraduate students in the Computer Science program in or recently completing the first course of the program. Participants prior programming, and specifically C programming, experience were collected in the pre-survey. Additionally participant prior LLM, and specifically ChatGPT, experience was collected in the pre-survey as well. Breakdowns of included participant pre-survey results are provided in~\autoref{sec:results}. 

\subsection{Instruments and Materials}
\subsubsection{Workstation}
\begin{figure*}[t]
    \centering
    \includegraphics[width=\textwidth]{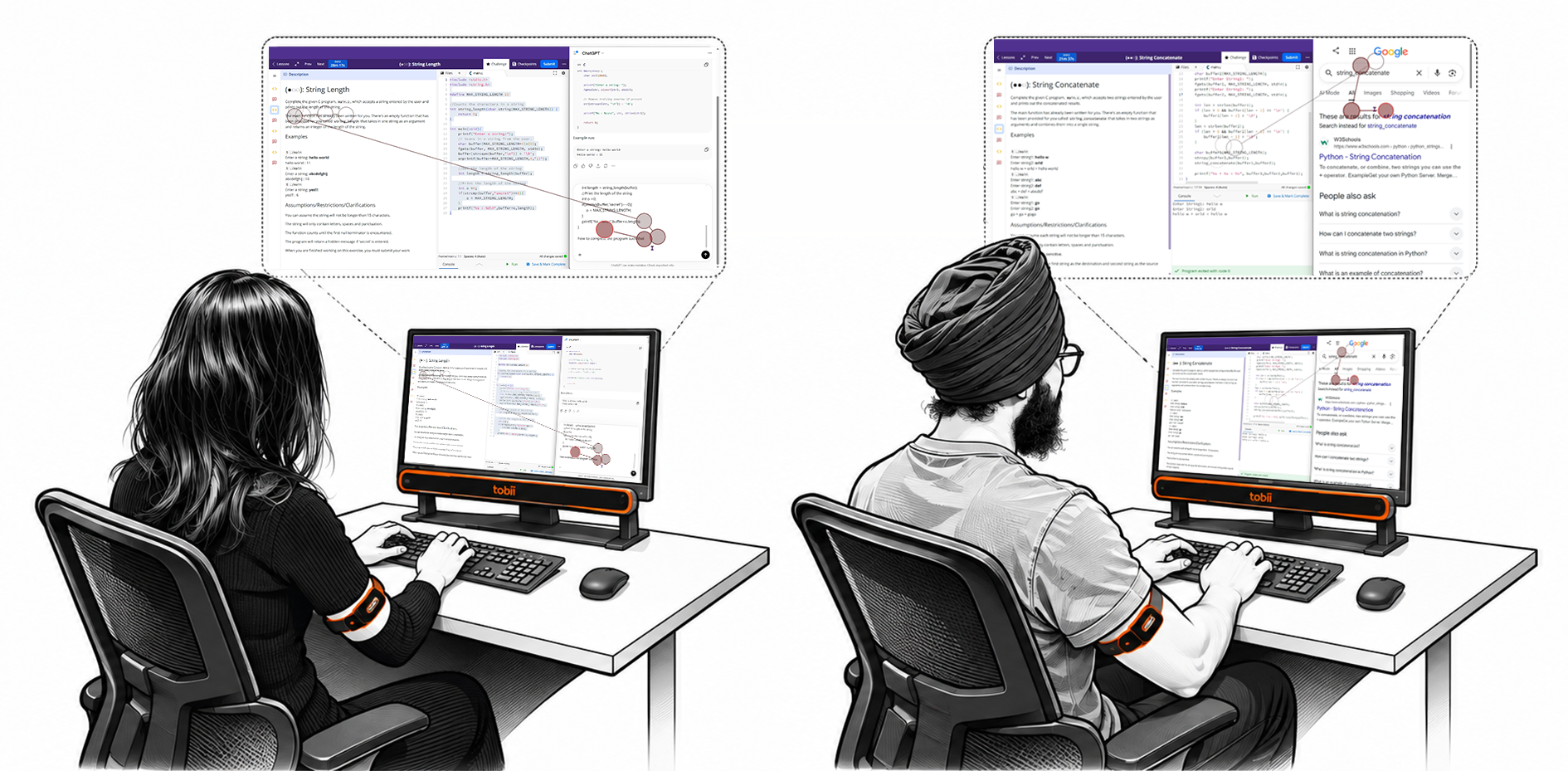}
    \caption{Experimental setup for the AI-Assisted (left) and No AI (right) conditions. Both used an eye tracker below the monitor and a heart rate sensor on the upper arm. Insets show the programming assessment beside ChatGPT or Google. Gaze markers are illustrative and were not visible to participants. Setup images were anonymised using GenAI, as disclosed in Section~\ref{sec:ai-disclosure}.}
    \Description{Two anonymised illustrations show participants at identical workstations. Each monitor displays a programming assessment and an assistance panel: ChatGPT on the left and Google on the right. Enlarged screen views illustrate gaze locations. An eye tracker sits beneath each monitor, and each participant wears an upper-arm heart rate sensor.}
    \label{fig:teaser}
\end{figure*}

The experiment was designed to simulate an average student completing a homework assessment for an introductory programming course. Both conditions used an identical physical setup: a single monitor, keyboard and mouse, with the monitor divided into two side-by-side panels. The right panel contained the condition-specific assistance application, ChatGPT or Google. The left panel contained the programming assessment in EdStem, which displayed the challenge instructions alongside a web-based development environment containing the initial scaffold code (~\autoref{fig:teaser}). Dividing the screen in this way allowed the eye tracking system to attribute gaze reliably to one application or the other. Monitor brightness and ambient lighting were held constant across sessions, and both assistance applications presented a light-background text interface, so that pupil diameter differences between conditions could not be attributed to display luminance. 

Participants in both conditions were told they could copy and paste any material from the assessment, including the full instructions and the scaffold code, into the application in the assistance panel (ChatGPT/Google). 

To mask variable ambient noise, an instrumental low-fi playlist was played in the laboratory space at a fixed volume for every participant. 

\subsubsection{Task Design}
The programming assessment was delivered through EdStem, the online educational platform used in participants' regular coursework, which ensured familiarity with the submission environment. The assessment was written in the C programming language, the language used in introductory courses at the participants' institution, and consisted of three programming questions of increasing difficulty. Each question presents a short C program with an incomplete function stub. Instructions for each question explicitly stated the function's input parameters and return type, provided example program output, and specified completion requirements such as maximum length of input variables. Submitted code was graded for completeness and accuracy of the specified function behaviour out of 5 possible points. 

\subsection{Measures}
\subsubsection{Cognitive Load}
Multiple physiological measurements of cognitive load were captured during the tasks. Gaze data was captured with a Tobii Pro Spectrum research-grade eye tracker. Pupil dilation has been shown to index cognitive load across several studies~\cite{Krejtz2018EyeGaze, Gavas2017EstimationDilation, Sibley2011PupilLearning}. Heart rate data was captured with the Polar Verity Sense, a consumer-grade heart rate tracker that has been used in human research as a less intrusive alternative to research-grade equipment~\cite{Schweizer2025Wrist-WornStudy.}. Participants were instructed to wear the sensor on the bicep in order to reduce any potential noise introduced by the movement of the sensor during typing. 

Perceived mental effort and difficulty were captured between each programming questions using Paas's Subjective Rating Scale of Mental Effort~\cite{Paas1992TrainingApproach, Paas1994InstructionalTasks, Paas2003CognitiveTheory}, comprising two 9-point Likert items on perceived mental load (low to high).

\subsubsection{Ownership}
Three post-task measures of authorship and ownership were captured. Authorship attribution was captured with a single 7-point Likert item developed by He et al.~\cite{He2025WhichCo-Creation}, ranging from sole authorship by the participant, to equal authorship, sole authorship by the AI/colleague. We utilised the Psychological Ownership Scale (POS), comprising four 5-point Likert questions (disagree to agree) assessing felt ownership over a target~\cite{VanDyne2004PsychologicalBehavior, Olckers2017MeasuringReview}, with the target adapted from the participant's organisation to the code they had submitted. Participants also reported, on a 100-point scale, what percentage of the submitted answers they considered their own rather than produced by GenAI or taken from the internet, adapted from Kosmyna et al.~\cite{Kosmyna2025YourTask} interview question. 

\subsubsection{Performance}
The three coding challenges were graded in EdStem out of 5 points each by automated unit tests validating program functionality and edge-case handling. Task performance is reported as the proportion of the available points achieved (total points out of 15 maximum points possible). 

Content retention was assessed by cued recall, using ten open-ended questions about the tasks completed during the experiment. To distinguish knowledge of this specific assessment from programming knowledge participants already held, the questions mixed general programming concepts with questions about specific edge cases and implementation details of the submitted solutions. The same cued recall instrument was administered twice, once immediately after the coding task, and again 48~hours later. The recruitment advertisement stated that a 48-hour follow-up would occur, and participants were reminded of it upon completing the initial recall task. Participants did not have access to their submitted code whilst responding to these questions.

Each response was scored independently by three markers using a three-point rubric ($0-2$) per question, giving a maximum of 20 points per assessment response. Markers were blind to the condition. A participant's content retention score is the mean of the three markers' totals, reported as the proportion of the available points achieved (mean markers' total out of 20 maximum points possible). Inter-rater reliability is reported in \autoref{sec:analysis}.

\subsection{Conditions}
\subsubsection{AI-Assisted}
Participants in the AI-Assisted condition were given access to ChatGPT-4.5. Settings were consistent across participants, with each participant logged into a different ChatGPT account, the memory feature turned off, and model fixed to GPT4.5. This ensures that each participant receives individualised responses not impacted by previous participants. The AI-Assisted experimental group were not given access to any other GenAI assistant tools or web search engines such as Google. 

\subsubsection{No AI}
The participants in the No AI control group did not have access to ChatGPT or any GenAI assistant tools. They were given access to the Google search engine, with all GenAI components disabled in the browser. Access to all major GenAI tool sites were blocked, and participants were instructed not to seek out GenAI tools through the search engine. Disabling Google's AI Overviews was a deliberate design decision to create a clear delineation between conventional web search and conversational GenAI assistance, accepting that this yields a different experience than students encounter in general use. It was acceptable for this group to not utilise any search tools. 

\subsection{Procedure}
Participants first completed a pre-survey capturing demographic information; recent sleep, recent caffeine intake, prior programming and LLM experience. Participants were briefed on the procedure and fitted with the heart rate sensor. 

After this was done, participants were seated at the workstation and the eye tracker was calibrated individually for each participant. To establish a physiological baseline under conditions comparable to the experimental task, participants were shown sample program code and completed a typing task copying the sample code for five minutes. 

Participants then had 30 minutes to complete the three programming problems, submitting each through EdStem as they finished it. EdStem displayed a time remaining timer and "Times Up" non-interrupting notification when time had elapsed. The Subjective Rating Scale of Mental Effort was administered between each question. 

If a participant appeared to make no progress for ten minutes, a predefined stuck procedure allowed the researcher to intervene. Intervention was limited to clarifying the task, offering conceptual prompts, or giving directional instructions. Researchers did not provide any code, algorithms, bug fixes or implementation guidance. This procedure was called upon four times, three of those times the participant did not finish all three coding challenges in the allotted time.

Once the assessment was complete or the time had elapsed, the tracking sensors were removed and recording stopped. Participant then completed the post-assessment survey containing ownership and authorship measures and the cued recall questions, without access to their submitted code. Following the post-survey, participants were debriefed and reminded of the follow up survey.

Forty-eight hours after the session, participants were sent an online survey containing the same recall questions. All participants completed the follow up survey. 

\subsection{Data Collection}
Self-reported data were collected through Qualtrics. Physiological data were recorded in real time during the session. Submitted code was captured automatically through EdStem. All data was stored securely in university-managed storage. 

\subsection{Analysis}
\label{sec:analysis}
All analyses were run in Python~3.13.7. Linear mixed models, ordinal logistic regression and Holm-Bonferroni corrections used Statsmodels~0.15.0; Welch's $t$ used SciPy~1.18.1.
The mixed ANOVA and Cronbach's $\alpha$ used pingouin~0.6.1. 
The Type~III $F$ tests, Satterthwaite degrees of freedom, $R^2$ and estimated marginal means for the mixed models were computed following the lmerTest algorithm~\citep{Kuznetsova2017LmerTestModels}.
Physiological signals were pre-processed with NeuroKit2~0.2.13~\citep{Makowski2021NeuroKit2:Processing}. We report Hedges' $g$ for two-group comparisons, and $\eta^2_p$ for omnibus effects, with 95\% confidence intervals. 

Group comparisons for assessment performance, retention immediately after, 48-hours after submission, self-attributed ownership and Psychological Ownership Scale scores used Welch's $t$-tests, which do not assume equal variance across groups. 

Adhering to recommendations on the disclosure of analytic choices~\cite{Simmons2011False-positiveSignificant}, we report all covariates considered. Reduced sleep and elevated caffeine intake in the 24 hours before the session were included as fixed effects in the physiological mixed models, since both plausibly affect pupil diameter and vagal heart rate variability (HRV). Prior programming experience, in any language and in C specifically, was examined as a covariate on performance and retention as participants may plausibly impact their performance on C based programming challenges and knowledge questions. Participants' prior Large Language Model (LLM) and specifically ChatGPT experience was examined as a covariate on ownership, since familiarity with LLMs could plausibly influence their relationship of ownership over generated content. No covariate was significant so all primary analyses are reported without covariates. 

Time on task was deliberately \emph{not} entered as a covariate: it is measured after random assignment and is itself shaped by the tool (Table~\ref{tab:time}), adjusting for it would condition on a post-treatment variable and estimate a direct effect net of any pathway through time, rather than the total effect of the tool that the research questions ask about~\citep{rosenbaum1984TheTreatment, montgomery2018HowIt}; ANCOVA assumes that the covariate is independent of group~\citep{miller2001MisunderstandingCovariance}. Time on task is therefore reported descriptively and treated as a candidate mechanism in \autoref{sec:discussion}.

\subsubsection{Performance}
The coding assessment consisted of three programming challenges scored out of 5 points each via automated unit tests in EdStem testing the described program requirements and edge cases (grade $=\text{points}/15$). Groups were compared using Welch's t-test, the recommended test for two independent groups~\cite{Delacre2017WhyT-Test}, with Hedges'~$g$~\cite{Lakens2013CalculatingANOVAs}.

An open-ended 10-question quiz was administered immediately after the session and again 48~hours later, and each question was scored by 3 independent raters using a three-point rubric ($0-2$). Inter-rater reliability was assessed using Krippendorff's $\alpha$~\cite{Krippendorff2011ComputingAlpha-Reliability} for ordinal data. Reliability was high overall ($\alpha=.94$) and remained high per question ($\alpha=.80$--$.99$)~\cite{krippendorff2004MeasuringData}. Each quiz used the same two-group tests as the programming challenge performance. The time effect was analysed with a $\text{group}\times\text{time}$ mixed ANOVA~\cite{Maxwell2004DESIGNINGEdition} where interaction tested potential differential forgetting. 

\subsubsection{Cognitive Load}
The primary measure is the 9-point self-reported scale of mental effort and difficulty collected after each task~\cite{Paas1992TrainingApproach,Paas2003CognitiveTheory}, treated as interval data. 

The physiological channels are supplemental and were measured where usable. An interval contributes pupil or heart rate data only if the task lasted at least 60 seconds~\cite{Shaffer2017AnNorms} and passes its quality gate~\cite{vanderWel2018PupilReview} (\autoref{tab:physioresponses}). Otherwise that task's pupil or heart rate value for that interval is treated as missing, while the session's self-reported cognitive load, performance, retention and ownership data are retained. 

Task pupillary response~\cite{Beatty1982Task-evokedResources,vanderWel2018PupilReview} was recorded with a Tobii Pro Spectrum and subtractive baseline correction~\cite{Mathot2018SafeData}. The Tobii Pro Lab whole-fixation mean pupil diameter (I-VT filter) was used if $\geq 20\%$ of the interval's duration is in fixations~\cite{Kret2018PreprocessingCode}. Task heart rate response was recorded with a Polar Verity Sense worn on the bicep, corrected with the artefact detection algorithm~\cite{Lipponen2019AClassification, Tarvainen2014KubiosSoftware} in NeuroKit2~\cite{Makowski2021NeuroKit2:Processing}. We compute ln RMSSD, the recommended HRV index for short segments~\cite{Shaffer2017AnNorms, Hjortskov2004TheWork}. An interval is only used if $\leq 25\%$ of successive pulse intervals differ by more than $20\%$~\cite{Malik1996HeartUse}. 

\begin{table}[ht]
    \centering
    \caption{Physiological intervals retained after quality gates.}
    \begin{tabular}{lcc}
    \toprule
        Interval & HRV Interval $n$ (\%)& Pupil Interval $n$ (\%) \\
    \midrule
        Baseline & 13 (25\%) & 47 (87\%) \\
        Task 1   & 19 (36\%) & 43 (80\%) \\
        Task 2   & 18 (35\%) & 42 (79\%) \\
        Task 3   & 16 (31\%) & 42 (81\%) \\
    \bottomrule
    \end{tabular}
    \label{tab:physioresponses}
\end{table}

Each load index (Paas effort, Paas difficulty, pupil~$\Delta$, ln RMSSD) was modelled separately with a linear mixed model, $y \sim \text{task} \times \text{tool}$,
with a random intercept per participant (Maximum Likelihood (ML) estimation). Mixed models were used because they accommodate the unbalanced data produced by the physiological quality gates. 
The confirmatory test for each index was a likelihood-ratio test of the tool$\,\times\,$task interaction against the additive model, fitted by maximum likelihood. The research question examined whether the condition changes the load trajectory across the assessment, which is a single model-comparison question per index. Per-task contrasts are reported as descriptive follow-up rather than as separate confirmatory tests. 

Fixed effects are reported from the REML fit as omnibus Type~III $F$ tests with sum-to-zero contrasts and Satterthwaite-approximated denominator degrees of freedom~\citep{Satterthwaite1946AnComponents}, with partial $\eta^2$ as the per-effect size for each term. Estimated marginal means (EMM) with Satterthwaite-$t$ 95\% CIs were computed on the balanced task$\,\times\,$tool grid. Model fit is summarised by marginal and conditional $R^2$~\citep{Nakagawa2013AModels}. Holm--Bonferroni correction was applied across the four confirmatory interaction tests. 

\subsubsection{Ownership}
The 7-point authorship attribution item~\cite{He2025WhichCo-Creation} was assessed via proportional-odds ordinal logistic regression on group. The percentage ownership slider ($0-100$)~\cite{Kosmyna2025YourTask} and the four-item Psychological Ownership Scale~\cite{VanDyne2004PsychologicalBehavior} were assessed with Welch's $t$-tests. Cronbach's $\alpha$~\cite{Tavakol2011MakingAlpha} was also calculated for the four scale items. 

All $p$ values are corrected within the three pre-specified test families above. Performance: the three Welch tests. Cognitive load: the four interaction tests across each cognitive channel index. Ownership: the three scale tests. Each family is corrected using Holm-Bonferroni correction procedure~\cite{Holm1979AProcedure}, which controls family-wise error rate. Uncorrected and Holm-corrected $p$ values are reported. 

\section{Results}
\label{sec:results}
Of the $N=59$ participants who completed the session, $n=55$ were retained for analysis (No AI $n=26$, AI-Assisted $n=29$). Four were excluded for protocol reasons: one participant allocated to the AI-Assisted condition did not use the assistance at any point, and three did not attempt Task~3 (two from No AI and one from AI-Assisted). 

~\autoref{tab:demographics} reports the composition of the analysed sample. Males comprised $78\%$ ($n=43$) of participants and females $22\%$ ($n=12$), consistent with the gender composition of Australian computer science programs~\cite{UniversityResources}. Few participants reported sleeping less than usual ($n=7$, $13\%$) or consuming more caffeine than usual ($n=3$, $5\%$) in the 24 hours prior to the session. All $n=55$ participants reported up to two years of experience with the C programming language specifically, and most ($n=39$,$71\%$) reported up to two years of programming experience in general, indicating a novice population. Most also reported no more than a moderate amount of experience with ChatGPT ($n=40$, $73\%$), and $80\%$ ($n=44$) indicating up to a moderate amount of experience with LLMs in general. 

\begin{table}[ht]
    \centering
    \caption{Included participants ($N=55$).}
    \label{tab:demographics}
    \begin{tabular}{lc}
        \toprule
          Group  & $n$ (\%) \\
        \midrule
          AI-Assisted & 29 (53\%) \\
          No AI & 26 (47\%)\\
        \multicolumn{2}{l}{\textbf{Gender}} \\
          Male & 43 (78\%) \\
          Female & 12 (22\%) \\
        \multicolumn{2}{l}{\textbf{Slept less than usual (24\,h)}} \\
          No & 48 (87\%) \\
          Yes & 7 (13\%) \\
        \multicolumn{2}{l}{\textbf{More caffeine than usual (24\,h)}} \\
          No & 52 (95\%) \\
          Yes & 3 (5\%) \\
        \multicolumn{2}{l}{\textbf{Programming experience (any language)}} \\
          No experience & 4 (7\%) \\
          1 Year & 22 (40\%) \\
          2 Years & 13 (24\%) \\
          3 Years & 11 (20\%) \\
          4 Years & 3 (5\%) \\
          5+ Years & 2 (4\%) \\
        \multicolumn{2}{l}{\textbf{C programming experience}} \\
          No experience & 5 (9\%) \\
          0-1 Year & 48 (87\%) \\
          1-2 Years & 2 (4\%) \\
        \multicolumn{2}{l}{\textbf{LLM experience}} \\
          None at all & 18 (33\%) \\
          A little & 6 (11\%) \\
          A moderate amount & 20 (36\%) \\
          A lot & 7 (13\%) \\
          A great deal & 4 (7\%) \\
        \multicolumn{2}{l}{\textbf{ChatGPT experience}} \\
          A little & 13 (24\%) \\
          A moderate amount & 27 (49\%) \\
          A lot & 11 (20\%) \\
          A great deal & 4 (7\%) \\
        \bottomrule
    \end{tabular}
\end{table}

\subsection{Performance}
\begin{figure}[ht]
    \centering    
    \includegraphics[width=0.75\columnwidth]{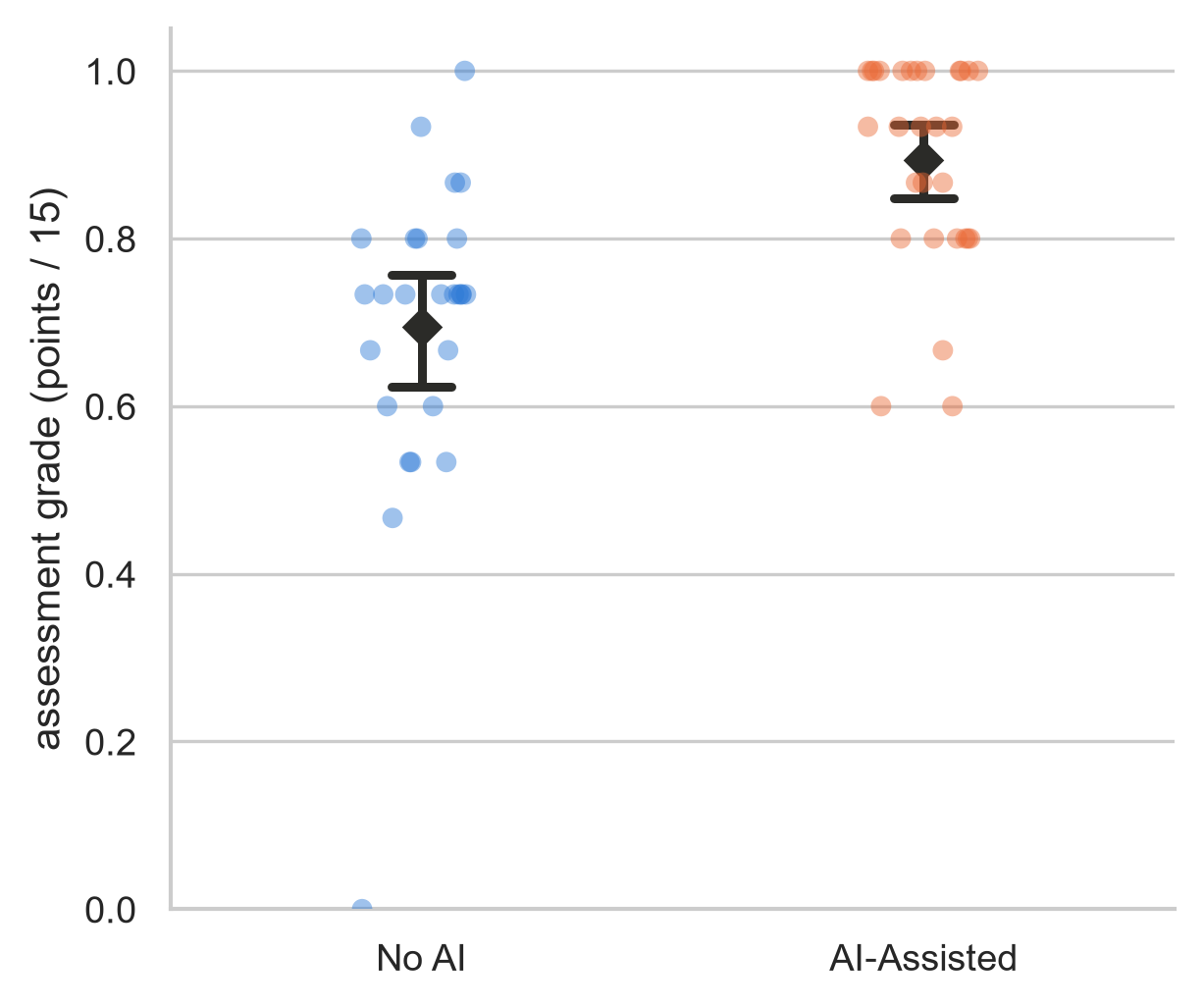}
    \caption{Task performance by assistance: individual grades, means and 95\% CIs.}
    \Description{Strip plot of assessment grades for the No AI and AI-Assisted groups with group means and 95 percent confidence intervals overlaid. The AI-Assisted group's grades cluster near the top of the scale, visibly above the more dispersed No AI group.}
    \label{fig:performance}
\end{figure}

The AI-Assisted group scored substantially higher on the coding assessment (AI-Assisted $M = 0.89$, $SD = 0.12$) than the No AI group ($M = 0.69$, $SD = 0.19$), Welch's $t(42.0)=-4.57$, $p<.001$, Hedges' $g =-1.24$, $p_\text{Holm}<.001$, shown in \autoref{fig:performance}. This significance survives Holm correction. 

\subsection{Time on Task}
Task durations derived from the Tobii interval markers (Table~\ref{tab:time}) show similar time on Task~1 in both conditions (No AI $M = 3.9$min, AI-Assisted $M = 4.0$min), diverging as difficulty increased. On Task~3 the No AI group worked nearly twice as long ($M = 11.4$min) as the AI-Assisted group ($M=5.9$min). Across the full assessment, the No AI participants spent $M = 24.1$min, while AI-Assisted participants took $M=16.4$min, sustaining their higher reported effort over a longer period (\autoref{sec:cog-load}). Three task intervals fell below the 60-second physiological gate (all in the AI-Assisted group) and are included in these descriptives. One AI-Assisted participant's Tobii recording failed to capture interval start and end times for any task, resulting in $n=28$ for the AI-Assisted descriptive analysis. These observations are descriptive only, no confirmatory statistical comparisons of time on task were conducted.

\begin{table}[ht]
    \centering
    \caption{Time on task in minutes, mean ($SD$) and median per task and tool including intervals under the 60s physiological gate. The Total row covers all three tasks.}
    \label{tab:time}
    \begin{tabular}{lcccccc}
        \toprule
         & \multicolumn{3}{c}{No AI} & \multicolumn{3}{c}{AI-Assisted} \\
        Task & $M$ ($SD$) & Mdn & $n$ & $M$ ($SD$) & Mdn & $n$ \\
        \midrule
        Task~1 & 3.9 (2.0) & 3.8 & 26 & 4.0 (2.8) & 3.3 & 28 \\
        Task~2 & 8.8 (5.7) & 6.4 & 26 & 6.4 (5.4) & 4.4 & 28 \\
        Task~3 & 11.4 (5.4) & 9.2 & 26 & 5.9 (4.3) & 5.5 & 28 \\
        Total & 24.1 (6.8) & 25.2 & 26 & 16.4 (8.7) & 16.4 & 28 \\
        \bottomrule
    \end{tabular}
\end{table}

\subsubsection{Content retention}
The No AI group scored higher on the retention quiz at both time points~(\autoref{tab:t-test}). Immediately after submission, the No AI participants~($M=0.53$, $SD=0.15$) recalled significantly more than the AI-Assisted group ($M=0.41$, $SD=0.17$) as confirmed by a Welch's $t$-test ($t(53.0)=2.68$, $p=.010$, $g=0.71$, $p_{\text{Holm}} = .011$). The same pattern was observed 48~hours later ($M=0.52$, $SD=0.16$ vs. $M=0.39$, $SD=0.19$ for No AI and AI-Assisted respectively), also significant ($t(52.9)=2.91$, $p=.005$, $g=0.77$, $p_{\text{Holm}} = .011$). Both findings survive Holm correction. 

\begin{table*}[ht]
    \centering\small
    \caption{Two-group comparisons (Welch's $t$). Effect size is Hedges' $g$.}
        \label{tab:t-test}
        \begin{tabular}{llcccccc}
        \toprule
        Family & Assessment & No AI $M(SD)$ & AI-Assisted $M(SD)$ & $t$ & $p$ & $p_{\text{Holm}}$ & $g$ \\
        \midrule
        Performance &Task performance & 0.69 (0.19) & 0.89 (0.12) & $t(42.0)=-4.57$ & <.001 & <.001 &-1.24 \\
        &Immediate retention & 0.53 (0.15) & 0.41 (0.17) & $t(53.0)=2.68$ & .010 & .011& 0.71 \\
        &48\,h retention & 0.52 (0.16) & 0.39 (0.19) & $t(52.9)=2.91$ & .005 & .011 & 0.77 \\
        \addlinespace
        
        Ownership&SAO 	& 80.81(27.48) & 44.59(38.68) &	$t(50.5) = 4.03$ & $<0.001$ & $<0.001$ & 1.05 \\
        &POS     & 4.30(0.85)   & 2.92(1.48)   &	$t(45.5) = 4.28$ & $<0.001$ & $<0.001$ & 1.11 \\
        \bottomrule
    \end{tabular}
\end{table*}

\begin{figure}[ht]
    \centering
    \includegraphics[width=0.75\columnwidth]{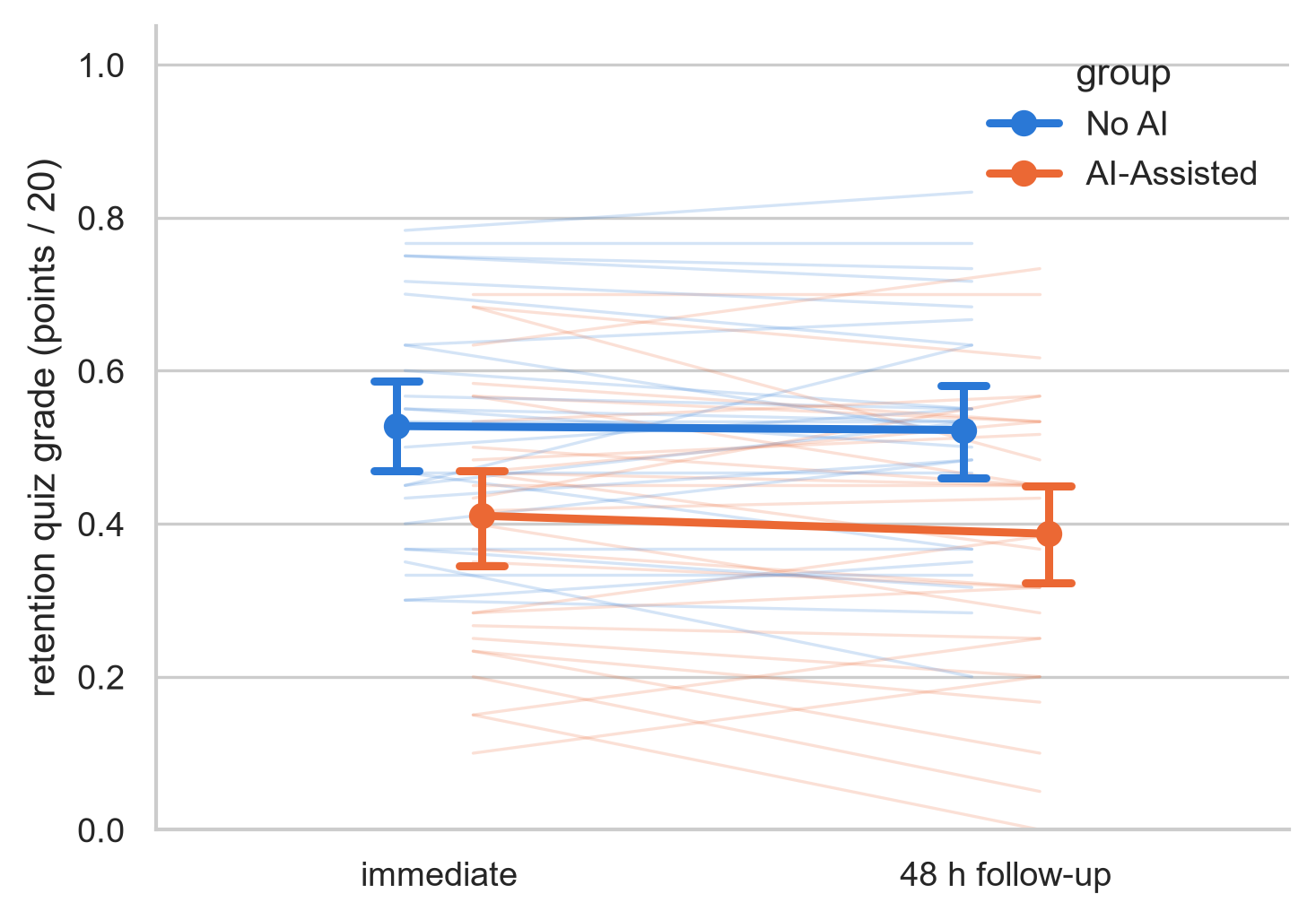}
    \caption{Retention grade by assistance at both time points, with individual trajectories.}
    \Description{Retention quiz grades at the immediate and 48-hour time points for the No AI and AI-Assisted groups, with thin lines showing each participant's trajectory and group means with 95 percent confidence intervals overlaid. The No AI group sits above the AI-Assisted group at both time points and the two group lines stay roughly parallel, indicating no differential forgetting.}
    \label{fig:retention}
\end{figure}

A mixed ANOVA confirmed the group effect ($F(1, 53) = 8.14$, $p = .006$, $\eta^2_p = 0.13$) with no main effect of time ($p = .169$) and, critically, no $\text{group}\times\text{time}$ interaction ($p = .391$). The gap in retention of the AI-Assisted group is established by the end of the session and remains the same 48~hours later, indicating there is no difference in the rate of forgetting between the groups (\autoref{fig:retention}).

\begin{figure}[ht]
    \centering
    \includegraphics[width=0.75\columnwidth]{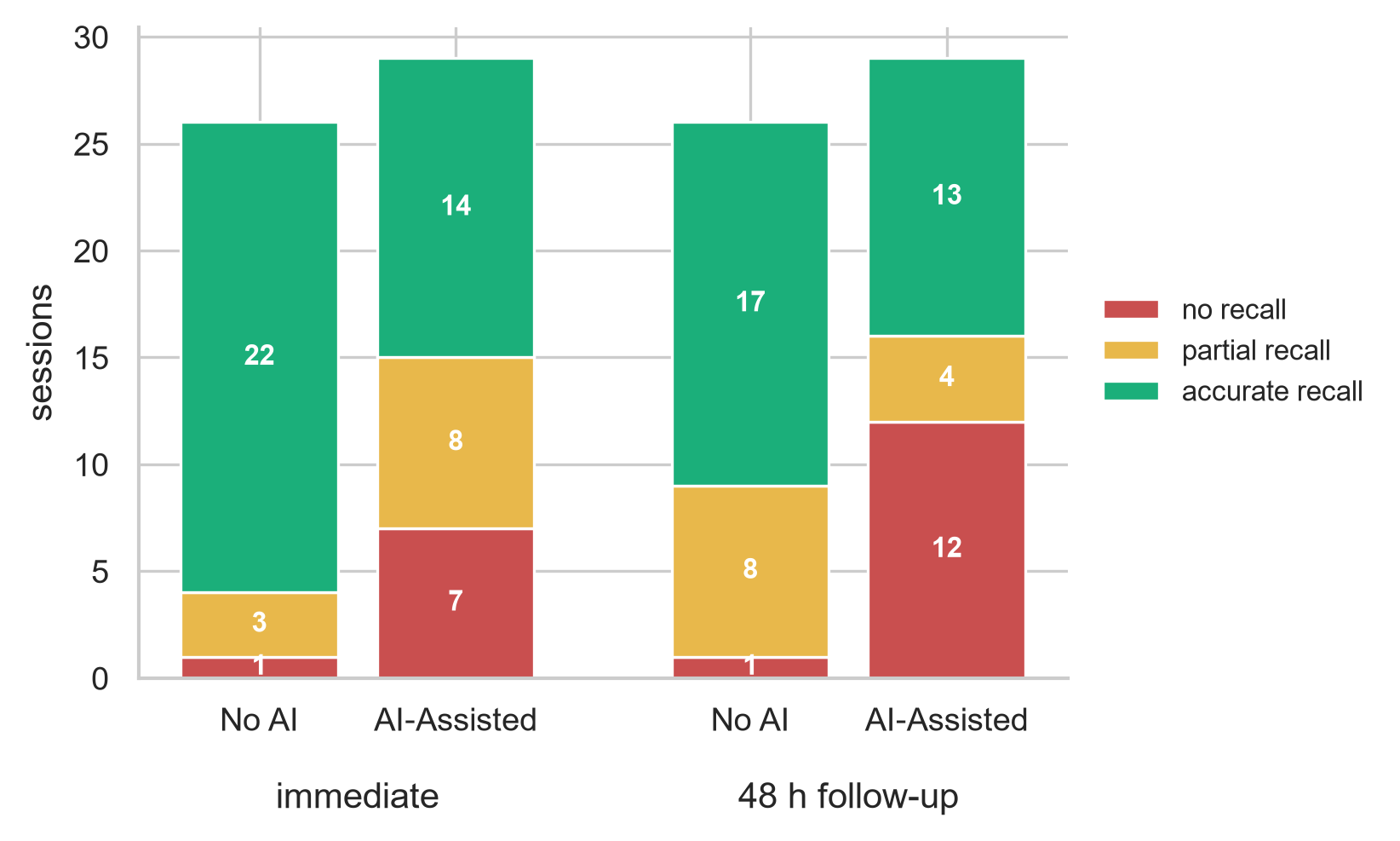}
    \caption{Code-line recall (retention quiz Q1) by tool and time point: counts of no, partial and accurate recall.}
    \Description{Stacked bar chart with four bars: the No AI and AI-Assisted groups at the immediate quiz and at the 48-hour follow-up. Each bar is divided into sessions with no recall, partial recall and accurate recall of a line of code from the participant's own submission. Accurate recall is more common in the No AI group and no recall is more common in the AI-Assisted group at both time points.}
    \label{fig:recall}
\end{figure}

The first question of the content retention quiz asked participants to recall a line of code from their own submission (0 = no recall, 1 = partial recall, 2 = accurate recall). Immediately after the session, $85\%$ of No AI participants ($n=22$) accurately recalled a line versus $48\%$ of AI-Assisted participants ($n=14$). Only $4\%$ ($n=1$) of No AI participants were unable to recall any line compared to $24\%$ ($n=7$) of AI-Assisted participants. At the 48-hour follow-up, accurate recall was $65\%$ ($n=17$) for the No AI participants and $45\%$ ($n=13$) for AI-Assisted participants, and the proportion unable to recall any line of code remained $4\%$ ($n=1$) for the No AI participants versus $41\%$ ($n=12$) AI-Assisted participants. These observations are descriptive only, no statistical comparisons of individual question performances were conducted (\autoref{fig:recall}).

\subsection{Cognitive Load}
\label{sec:cog-load}
For each index, the likelihood-ratio test of the tool$\,\times\,$task interaction was conducted (Table~\ref{tab:load-lr}), testing whether the conditions changed the load trajectory across the assessment. A significant tool$\times$task interaction was identified for self-reported mental effort. Omnibus Type~III $F$ tests then decomposed each model into Group, Task and Group$\,\times\,$Task effects with partial $\eta^2$ (Table~\ref{tab:load-omnibus}), and the per-task fixed effects, random-effect variances and $R^2$ (Table~\ref{tab:load-lmm}), describing how participants progressed through the tasks. In the coefficient table the reference level is Task~1 in the AI-Assisted group: the Task~2 and Task~3 rows give that group's change from Task~1, the No AI row gives the group difference at Task~1, and the interaction rows give the additional change in the No AI group on the later tasks.

Physiological quality gate definitions are included in \autoref{sec:analysis} and counts of surviving physiological responses are available in \autoref{tab:physioresponses}.

\begin{table}[ht]
    \centering
    \caption{Confirmatory likelihood-ratio test of the tool$\,\times\,$task interaction (full vs. additive model) per cognitive-load index.}
    \label{tab:load-lr}
    \begin{tabular}{lccccc}
        \toprule
        Index               & Obs.& $n$ & $\chi^2(2)$ & $p$ & $p_{\text{Holm}}$ \\
        \midrule        
        Mental Effort       & 165 & 55 & $8.88$ & $.012$ & $.047$ \\
        Difficulty          & 165 & 55 & $2.44$ & $.296$ & $.591$ \\
        Pupil $\Delta$ (mm) & 118 & 44 & $4.56$ & $.103$ & $.308$ \\
        ln RMSSD            & 53  & 24 & $0.32$ & $.853$ & $.853$ \\

        \bottomrule
    \end{tabular}
\end{table}

\begin{figure}[ht]
    \centering
    \includegraphics[width=0.75\columnwidth]{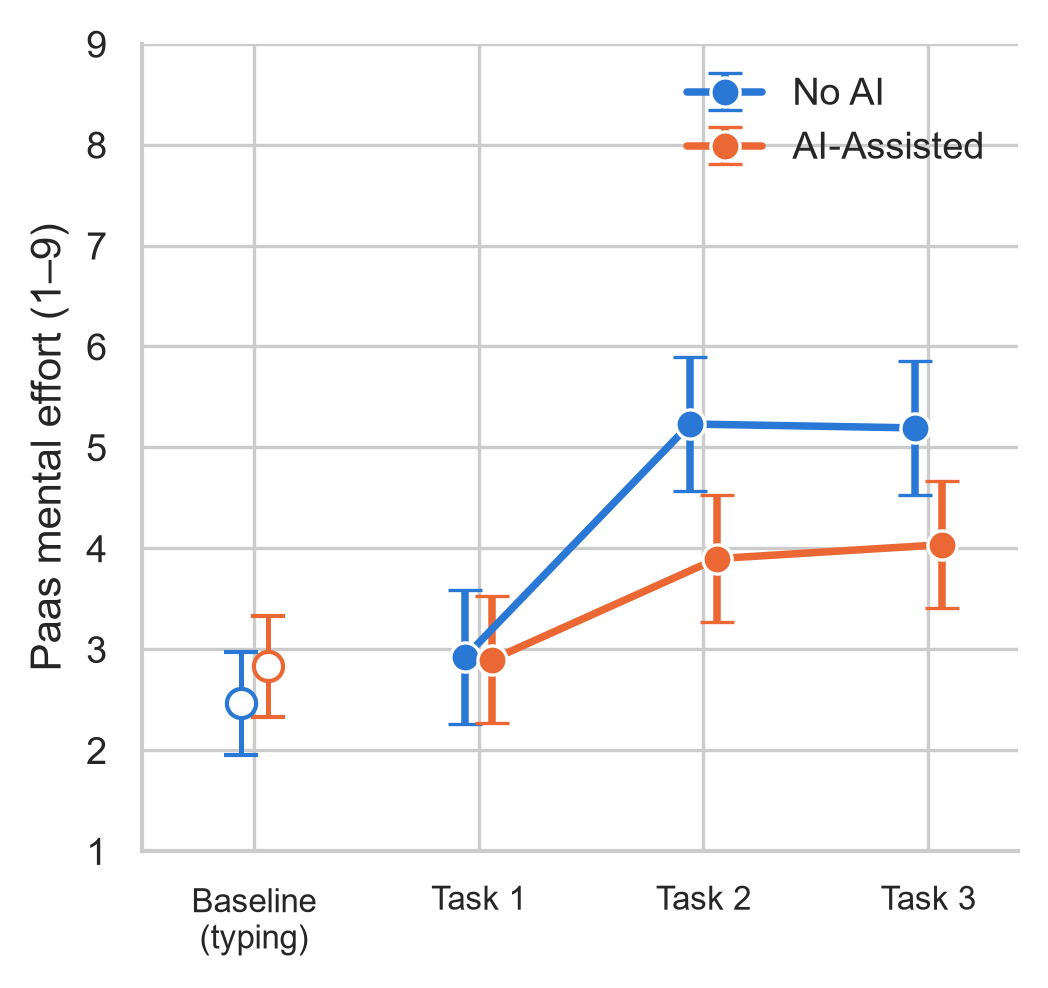}
    \caption{Self-reported mental effort EMM per task with 95\% confidence intervals.}
    \Description{Line chart showing mean mental effort ratings on a 1 to 9 scale for the No AI and AI-Assisted groups across a typing baseline and three tasks, with error bars indicating variability. Mental effort increases from baseline to later tasks in both groups. Ratings are similar in Task 1. In Tasks 2 and 3, participants report higher mental effort in the No AI group than in the AI-Assisted group. The largest difference occurs during Task 2. Overall, GenAI assistance is associated with lower perceived mental effort during the more demanding tasks.}
    \label{fig:effort_load}
\end{figure}

There was a significant tool$\times$task interaction for self-reported mental effort ($\chi^2(2)=8.88$, $p=.012$, $p_\text{Holm}=.047$) and survives Holm correction. Follow-up analysis revealed a significant group$\times$task interaction ($F(2,106.0)=4.46$, $p=.014$). Specifically, the two groups began the session with no significant difference on reported mental effort at Task~1 ($\beta=0.03$, 95\% CI$[-0.86,0.91]$, $p = .953$), indicating any divergence arose during the assessment rather than reflecting a prior group difference. As the tasks became harder, effort rose across both groups. A significant task $\times$ group effect was observed as mental effort increased further in the No AI group at both Task~2 ($\beta=1.31$, 95\% CI$[0.39,2.22]$ ,$p=.005$) and Task~3 ($\beta=1.13$, 95\% CI$[0.22,2.05]$, $p=.015$). See~\autoref{fig:effort_load}.

\begin{figure}
    \centering
    \includegraphics[width=0.75\columnwidth]{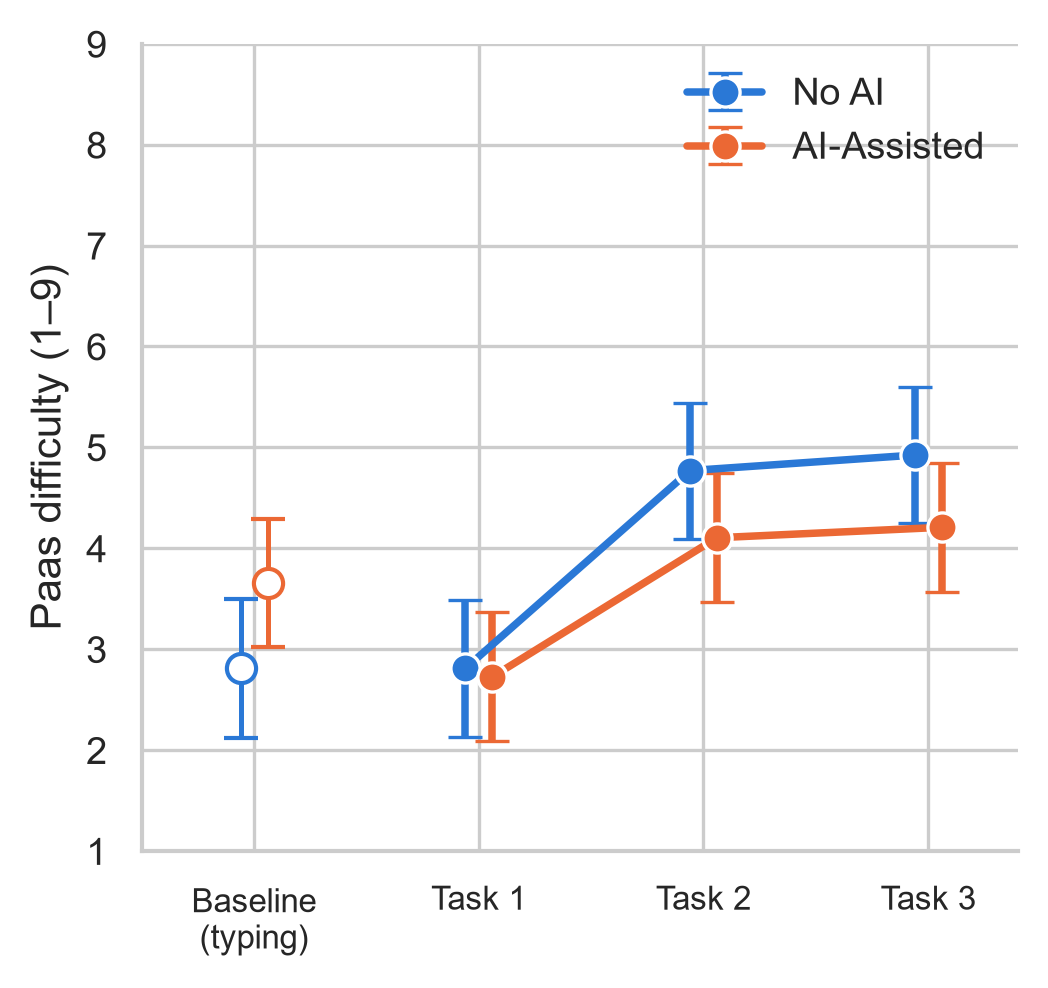}
    \caption{Self-reported task difficulty EMM per task with 95\% confidence intervals.}
    \Description{Line chart showing mean difficulty ratings on a 1 to 9 scale for No AI and AI-assisted group across a typing baseline and three tasks, with error bars indicating variability. Difficulty ratings increase from baseline to Tasks 2 and 3 in both group. At baseline, the AI-assisted group is rated slightly more difficult than the No AI group. Ratings are similar in Task 1. In Tasks 2 and 3, perceived difficulty is higher in the No AI group than in the AI-assisted group. Overall, GenAI assistance is associated with lower perceived task difficulty for the later tasks.}
    \label{fig:difficulty_load}
\end{figure}

There was no significant tool$\times$task interaction for self-reported difficulty ($\chi^2(2) = 2.44$, $p = .296$, $p_\text{Holm}=.591$) and remains insignificant after Holm correction. However, follow-up analysis does show a significant main effect of task ($F(2,106.0)=38.71$, $p<.001$). That effect held across all three tasks as both groups rated the subsequent tasks more difficult ($\beta=1.38$, 95\% CI$[0.78,1.98]$, $p<.001$; $\beta=1.48$, 95\% CI$[0.88,2.09]$, $p<.001$). See \autoref{fig:difficulty_load}

The baseline interval was a copy-typing task, chosen to place a realistic motor-plus-attention load on every participant so that task-level findings could not be attributed to the experimental setup alone. The self-reported cognitive load ratings collected after the baseline indicate it worked as intended. Participants rated the typing task at a level of mental effort comparable to Task~1 (No AI: baseline $M=2.46$ vs. Task~1 $M=2.92$; AI-Assisted: baseline $M=2.83$ vs. Task~1 $M=2.90$), and $49\%$ ($n=27$) of participants rated baseline effort at or above their Task~1 effort. For perceived difficulty, No AI participants rated the typing test and Task~1 identically (baseline $M=2.81$ vs. Task~1 $M=2.81$), whereas AI-Assisted participants rated the typing test as \emph{harder} than Task~1 (baseline $M=3.66$ vs. Task~1 $M=2.72$). This matters for reading the pupil deltas: a near-zero or negative delta does not indicate an unloaded task, only load comparable to or below an already-demanding reference. The deltas are therefore interpretable as change relative to the baseline anchor, as opposed to absolute load.

\begin{figure}
    \centering
    \includegraphics[width=0.75\columnwidth]{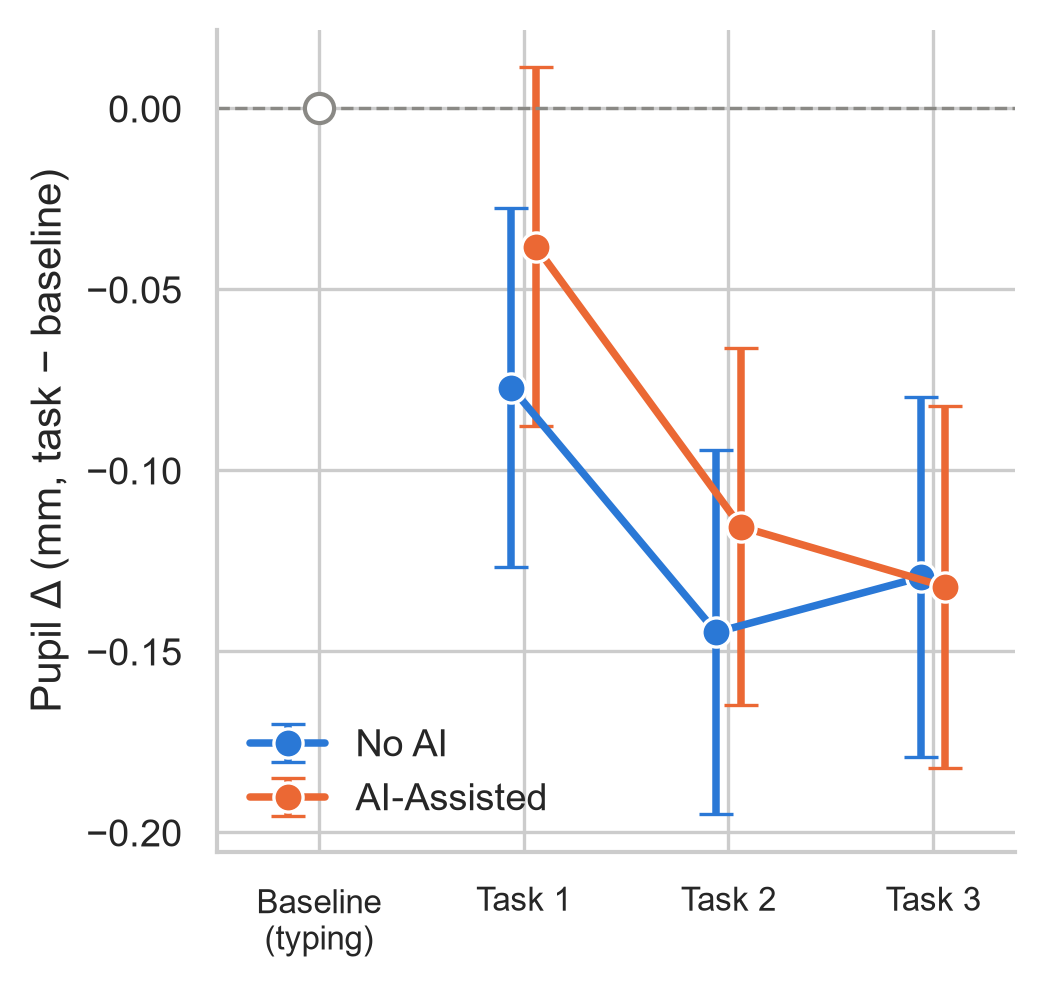}
    \caption{Pupil diameter delta (mm, task-baseline) EMM per task with 95\% confidence intervals.}
    \Description{Line chart showing mean change in pupil diameter relative to a typing baseline for the No AI and AI-Assisted groups across three tasks, with error bars indicating variability. Baseline pupil diameter change is zero. Both groups show pupil constriction (negative values) during all tasks. The AI-Assisted condition shows a smaller reduction in pupil diameter than the No AI condition in Task 1 and Task 2. By Task 3, the two groups converge. Overall, pupil diameter decreases from baseline during task performance, with larger reductions in the No AI condition.}
    \label{fig:pupil_load}
\end{figure}

There was no significant tool$\times$task interaction for pupil $\Delta$ ($\chi^2(2) = 4.56$, $p = .103$, $p_\text{Holm}=.308$) and remains insignificant after Holm correction. The follow-up analysis showed a significant main effect of task ($F(2,70.4)=32.72$, $p<.001$). Both groups showed significant reductions in pupil$\Delta$ (reduced pupil dilation) at Task~2 and Task~3 ($\beta=-0.08$, 95\% CI$[-0.11,-0.05]$, $p<.001$; $\beta=-0.09$, 95\% CI$[-0.12,-0.07]$ $p<.001$). A significant group$\times$task contrast was observed at Task~3$\times$No AI ($\beta=0.04$, 95\% CI$[0.002,0.08]$, $p=.039$), where the AI-Assisted participants exhibited a significantly larger negative pupil$\Delta$ (smaller pupil) in comparison to the No AI group. See~\autoref{fig:pupil_load}. Task level significance is exploratory and was not a part of our confirmatory testing. 

There was no significant tool$\times$task interaction for ln RMSSD ($\chi^2(2) = 0.32$, $p = .853$, $p_\text{Holm}=.853$) and remained insignificant after Holm correction. Follow-up omnibus $F$ tests also did not identify any significant group$\times$task effect ($F(2,27.8)=0.14$, $p=.868$).

\begin{table}[ht] 
    \centering 
    \caption{Omnibus Type~III $F$ tests per cognitive-load index.} 
    \label{tab:load-omnibus} 
    \begin{tabular}{llcccc} 
        \toprule 
            Index & Effect & $df$ & $F$ & $p$ & $\eta_p^2$ \\
        \midrule 
            Mental Effort & Group & 1, 53.0 & 5.14 & .027 & .088 \\ 
            & Task & 2, 106.0 & 33.28 & $<.001$ & .386 \\ 
            & G$\times$T & 2, 106.0 & 4.46 & .014 & .078 \\ 
            \addlinespace 
            Difficulty & Group & 1, 53.0 & 1.59 & .212 & .029 \\
            & Task & 2, 106.0 & 38.71 & $<.001$ & .422 \\ 
            & G$\times$T & 2, 106.0 & 1.19 & .309 & .022 \\ 
            \addlinespace 
            Pupil $\Delta$ & Group & 1, 41.2 & 0.44 & .512 & .010 \\ 
            & Task & 2, 70.4 & 32.73 & $<.001$ & .482 \\ 
            & G$\times$T & 2, 70.4 & 2.21 & .117 & .059 \\ 
            \addlinespace ln RMSSD & Group & 1, 22.6 & 0.01 & .927 & .000 \\ 
            & Task & 2, 27.8 & 0.30 & .743 & .021 \\ 
            & G$\times$T &  2, 27.8 & 0.14 & .868 & .010 \\ 
            \bottomrule 
    \end{tabular} 
\end{table}

\begin{table*}[ht]
    \centering
    \caption{Linear mixed models per cognitive-load index ($y \sim \text{task} \times \text{tool}$, random intercept per participant, ML): fixed-effect estimates, random-effect variances, and marginal / conditional $R^2$. Reference level: Task~1 in the AI-Assisted group.}
    \label{tab:load-lmm}
    \begin{tabular}{llccccc}
    \toprule
    Index & Predictor & $b$ & $SE$ & 95\% CI & $t$ & $p$ \\
    \midrule    
    Mental effort & Intercept (Task~1, AI-Assisted) & 2.90 & 0.31 & [2.29, 3.51] & 9.31 & $<.001$ \\
     & Task~2 (vs.\ Task~1) & 1.00 & 0.32 & [0.37, 1.63] & 3.12 & $= .002$ \\
     & Task~3 (vs.\ Task~1) & 1.14 & 0.32 & [0.51, 1.77] & 3.55 & $<.001$ \\
     & No AI (at Task~1) & 0.03 & 0.45 & [-0.86, 0.91] & 0.06 & $= .953$ \\
     & Task~2 $\times$ No AI & 1.31 & 0.47 & [0.39, 2.22] & 2.80 & $= .005$ \\
     & Task~3 $\times$ No AI & 1.13 & 0.47 & [0.22, 2.05] & 2.42 & $= .015$ \\
     & \emph{Participant intercept $\sigma^2_{u}$} & 1.313 & \multicolumn{2}{l}{$SD = 1.15$} & & \\
     & \emph{Residual $\sigma^2_{\varepsilon}$} & 1.492 & \multicolumn{2}{l}{$SD = 1.22$} & & \\
     & \multicolumn{6}{l}{\emph{$R^2_{\text{marginal}} = 0.235$, $R^2_{\text{conditional}} = 0.593$}} \\
    \addlinespace
    Difficulty & Intercept (Task~1, AI-Assisted) & 2.72 & 0.32 & [2.10, 3.34] & 8.62 & $<.001$ \\
     & Task~2 (vs.\ Task~1) & 1.38 & 0.31 & [0.78, 1.98] & 4.48 & $<.001$ \\
     & Task~3 (vs.\ Task~1) & 1.48 & 0.31 & [0.88, 2.09] & 4.82 & $<.001$ \\
     & No AI (at Task~1) & 0.08 & 0.46 & [-0.82, 0.98] & 0.18 & $= .856$ \\
     & Task~2 $\times$ No AI & 0.58 & 0.45 & [-0.30, 1.46] & 1.30 & $= .194$ \\
     & Task~3 $\times$ No AI & 0.63 & 0.45 & [-0.25, 1.51] & 1.41 & $= .158$ \\
     & \emph{Participant intercept $\sigma^2_{u}$} & 1.520 & \multicolumn{2}{l}{$SD = 1.23$} & & \\
     & \emph{Residual $\sigma^2_{\varepsilon}$} & 1.375 & \multicolumn{2}{l}{$SD = 1.17$} & & \\
     & \multicolumn{6}{l}{\emph{$R^2_{\text{marginal}} = 0.203$, $R^2_{\text{conditional}} = 0.622$}} \\
    \addlinespace
    Pupil $\Delta$ (mm) & Intercept (Task~1, AI-Assisted) & -0.04 & 0.024 & [-0.086, 0.009] & -1.58 & $= .113$ \\
     & Task~2 (vs.\ Task~1) & -0.08 & 0.01 & [-0.105, -0.050] & -5.55 & $<.001$ \\
     & Task~3 (vs.\ Task~1) & -0.09 & 0.01 & [-0.122, -0.066] & -6.59 & $<.001$ \\
     & No AI (at Task~1) & -0.04 & 0.03 & [-0.106, 0.028] & -1.14 & $= .254$ \\
     & Task~2 $\times$ No AI & 0.01 & 0.02 & [-0.030, 0.049] & 0.48 & $= .629$ \\
     & Task~3 $\times$ No AI & 0.04 & 0.02 & [0.002, 0.081] & 2.06 & $= .039$ \\
     & \emph{Participant intercept $\sigma^2_{u}$} & 0.0106 & \multicolumn{2}{l}{$SD = 0.103$} & & \\
     & \emph{Residual $\sigma^2_{\varepsilon}$} & 0.0019 & \multicolumn{2}{l}{$SD = 0.043$} & & \\
     & \multicolumn{6}{l}{\emph{$R^2_{\text{marginal}} = 0.099$, $R^2_{\text{conditional}} = 0.864$}} \\
    \addlinespace
    ln RMSSD & Intercept (Task~1, AI-Assisted) & 4.45 & 0.15 & [4.16, 4.73] & 30.58 & $<.001$ \\
     & Task~2 (vs.\ Task~1) & -0.12 & 0.11 & [-0.34, 0.10] & -1.04 & $= .298$ \\
     & Task~3 (vs.\ Task~1) & -0.05 & 0.12 & [-0.29, 0.18] & -0.45 & $= .654$ \\
     & No AI (at Task~1) & -0.07 & 0.22 & [-0.50, 0.36] & -0.32 & $= .748$ \\
     & Task~2 $\times$ No AI & 0.09 & 0.17 & [-0.24, 0.42] & 0.56 & $= .577$ \\
     & Task~3 $\times$ No AI & 0.06 & 0.18 & [-0.29, 0.41] & 0.33 & $= .738$ \\
     & \emph{Participant intercept $\sigma^2_{u}$} & 0.201 & \multicolumn{2}{l}{$SD = 0.45$} & & \\
     & \emph{Residual $\sigma^2_{\varepsilon}$} & 0.057 & \multicolumn{2}{l}{$SD = 0.24$} & & \\
     & \multicolumn{6}{l}{\emph{$R^2_{\text{marginal}} = 0.006$, $R^2_{\text{conditional}} = 0.782$}} \\

    \bottomrule
    \end{tabular}
\end{table*}

\subsection{Sense of Code Ownership}
\begin{figure}
    \centering
    \includegraphics[width=0.75\columnwidth]{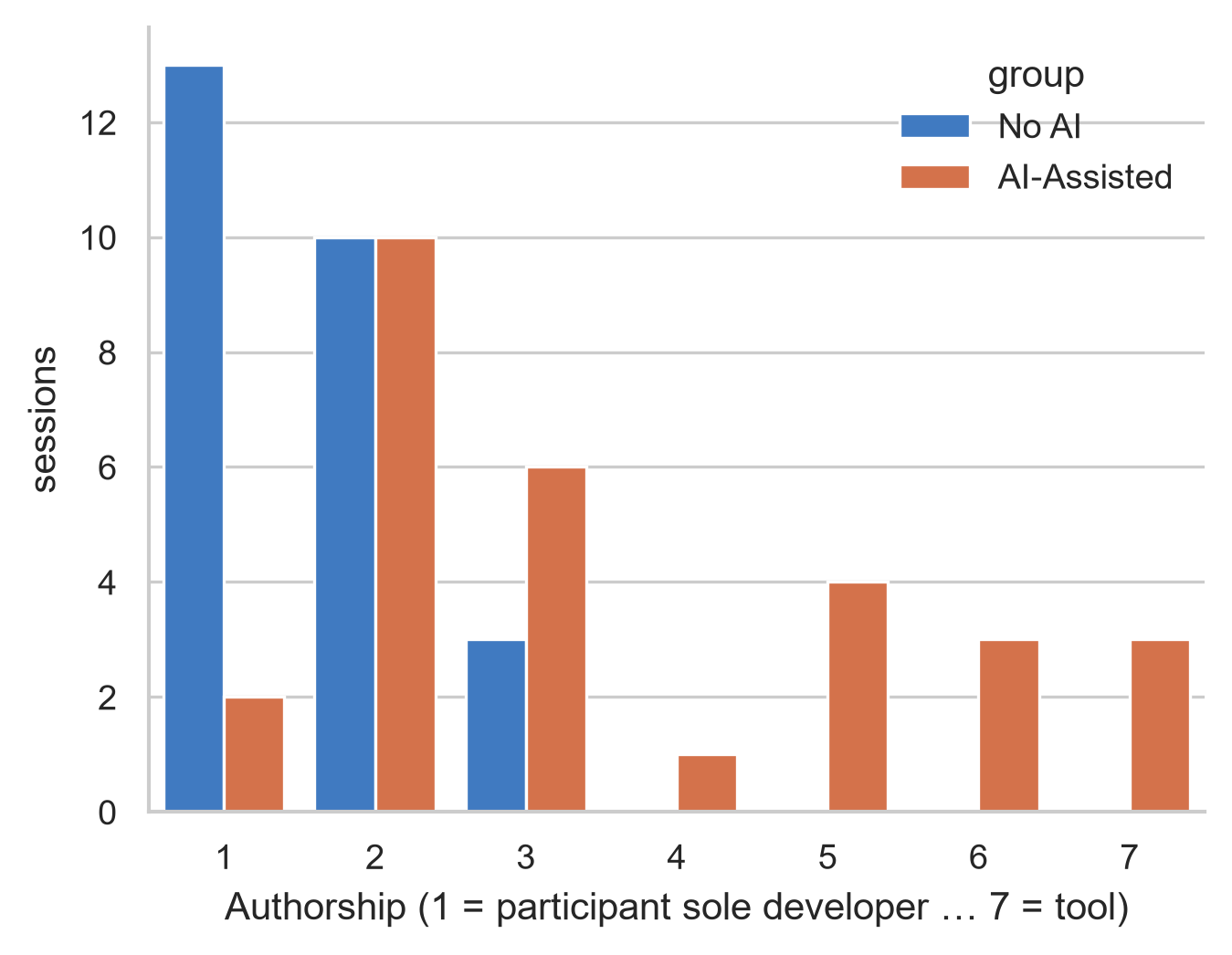}
    \caption{Authorship attribution distribution by group}
    \Description{Grouped histogram showing the number of sessions associated with authorship attribution ratings from 1 to 7, where 1 indicates the participant is viewed as the sole developer and 7 indicates the tool is viewed as the sole developer. In the No AI condition, ratings are concentrated at the participant end of the scale, with most sessions rated 1 or 2 and no ratings above 3. In the AI-Assisted condition, ratings are distributed across the full scale from 1 to 7. While ratings of 2 and 3 remain common, substantial numbers of sessions receive higher ratings between 4 and 7, indicating that participants attributed greater authorship to the AI when assistance was available. Overall, GenAI assistance shifts perceived authorship away from the participant and toward shared or tool-driven contributions.}
    \label{fig:authorship}
\end{figure}

AI-Assisted participants reported significantly lower ownership on all three measures, and all three effects survived Holm correction (\autoref{fig:authorship}). Authorship attribution was captured on a seven-point scale running from sole authorship by the participant to sole authorship by the tool, with intermediate points for acknowledgement, secondary authorship and equal co-authorship (Appendix~\ref{sec:Authorship}). AI-Assisted participants were significantly more likely to attribute co-authorship or primary authorship to ChatGPT than the No AI participants were to attribute authorship to Google ($\mathrm{OR} = 13.4$ $[4.0, 44.8]$, $p<.001$ $p_{\text{Holm}} <.001$). 

Half of the No AI group ($n=13$) claimed sole authorship, $n=10$ ($38\%$) would acknowledge Google without listing it as an author, and $n=3$ ($12\%$) noted they would list Google as a secondary author. In the AI-Assisted group, $n=2$ participants ($7\%$) claimed full sole authorship, $n=16$ ($55\%$) would acknowledge or list ChatGPT as a secondary author, and $n=11$ ($38\%$) would attribute ChatGPT as the primary author or co-author. 

\begin{figure}[ht]
    \centering
    \includegraphics[width=0.75\columnwidth]{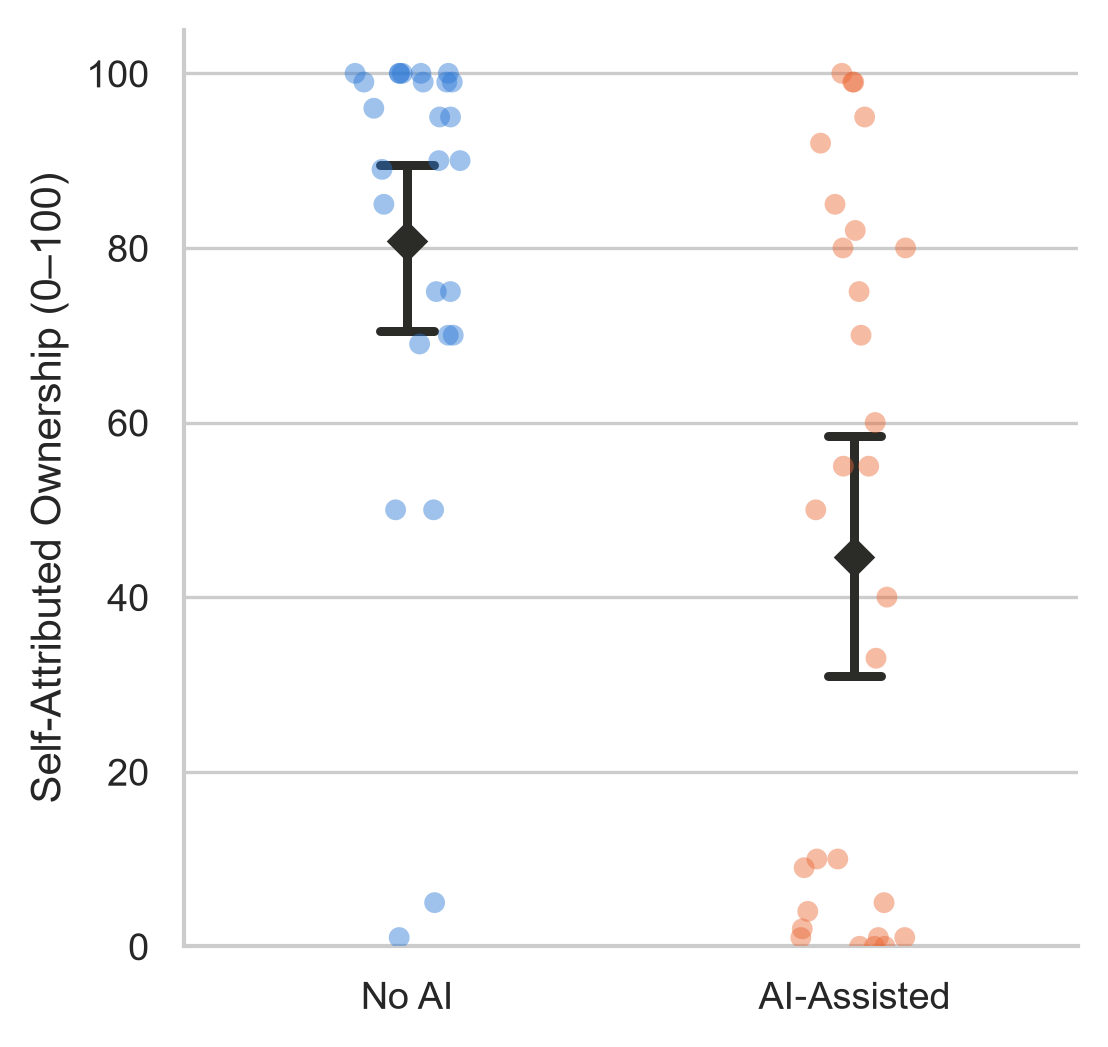}
    \caption{Self-attributed ownership percentage by group with 95\% CIs}
    \Description{Dot plot showing individual self-attributed ownership scores (0 to 100) for the No AI and AI-Assisted groups. Black diamonds indicate group means and error bars indicate variability. In the No AI condition, most ownership ratings cluster between 70 and 100, with a mean near 80, although a small number of observations fall close to zero. In the AI-Assisted condition, ratings are more widely distributed, spanning the entire scale from 0 to 100, with a lower mean near 45. Several observations in the AI-Assisted condition are close to zero, while others remain high. Overall, participants reported substantially lower and more variable ownership of their work when using GenAI assistance.}
    \label{fig:ownership_percentage}
\end{figure}

Self-attributed ownership (SAO), the percentage of submitted code that the participants considered their own, was lower in the AI-Assisted condition ($M = 44.6$) than the No AI condition ($M = 80.8$), $t(50.5) = 4.03$, $p<.001$, $p_{\text{Holm}} <.001$, $g = 1.05$. See~\autoref{fig:ownership_percentage}.

The final measure, Psychological Ownership Scale (POS), scores were also lower for the AI-Assisted condition ($M = 2.92$) than the No AI condition ($M = 4.30$), $t(45.5) = 4.28$, $p<.001$ $p_{\text{Holm}} <.001$, $g = 1.11$) See~\autoref{tab:t-test}. Cronbach's $\alpha = 0.94$.

\section{Discussion}
\label{sec:discussion}
The status of germane load is debated within Cognitive Load Theory (CLT); we follow Sweller and colleagues in treating germane load not as an additive third source of load but as the working memory resources devoted to dealing with intrinsic load, such that it does not add to the total load and instead describes where available capacity is directed~\cite{Sweller2011CognitiveTheory}.

\subsection{Performance Gains vs Retention Costs}
\textbf{RQ1a}
The AI-Assisted group significantly outperformed the No AI group on the assessment with an average grade of $89\%$ compared to $69\%$. Each coding challenge was an introductory level task, within the capability of the model to solve when given the full context. 

Several of the instances where AI-Assisted participants did not receive full marks involved participants who had not provided the full instructions or the scaffold code to the model. One AI-Assisted participant did not complete all three tasks. This participant attempted to use ChatGPT as a tutor, asking the model to respond with questions or next step hints, rather than solutions. The exchange became circular and self-contradicting and did not provide useful responses to the student, resulting in the student being unable to finish Task~2 before the timer had expired. This student was unable to attempt Task~3. Meanwhile, the two participants in the No AI group that were also unable to attempt Task~3 were among a small subset in the No AI group that attempted to treat Google search as an AI. We hypothesise that this behaviour is due to increasing reliance on GenAI tools, or even unfamiliarity with pre-GenAI search. However, with AI Summaries disabled, this yielded limited relevant results and reduced Google's usefulness as a search tool.

\textbf{RQ1c}
Cognitive offloading appears to come at a cost of retention of information. The No AI group outperformed the AI-Assisted group $53\%$ to $41\%$ immediately following the assessment and this same differential was maintained even 48~hours later ($52\%$, $39\%$). This indicates that both groups forgot at the same rate over those 48~hours, but the AI-Assisted group did not retain as much information during the tasks. Over half ($52\%$, $n=15$) of the AI-Assisted participants could not accurately recall a line of code they had submitted immediately after submission (no or partial recall), with one of those unable to recall simply responding "Absolutely not". Meanwhile $85\%$ ($n=22$) of No AI participants accurately recalled a line of code immediately after submitting. If AI assistance was only reducing extraneous cognitive load while retaining intrinsic load, CLT posits that AI-Assisted participants' germane load should have allocated working memory to develop new or update existing schema of the information that would have been available to them in the recall stages. 

\subsection{Interpreting Reduced Cognitive Load}
\textbf{RQ1b}
While both groups reported the mental effort of Task~1 similarly, the No AI participants reported significantly higher mental effort than the AI-Assisted participants for Task~2 and Task~3. Combined with the reported difficulty, participants in both groups acknowledged that the tasks were getting harder, but only No AI participants reported commensurate increases in mental effort. Taken together, it appears that the AI-Assisted group appears to be aware the problem difficulty is rising, however, they appear to remain disengaged from the content itself. This pattern is consistent with offloading the cognitive work to the tool~\cite{Stadler2024CognitiveInquiry}.

This pattern could be substantiated by time on task as both groups spent similar time on Task~1, again implying that the two groups began at similar states. However, the time spent on later tasks diverged, notably, the No AI participants spent nearly twice as long on Task~3 with three of the AI-Assisted participants completing Task~3 in under $60$ seconds. Overall, the No AI participants spent roughly 1.5$\times$ as long working through the coding challenges. This is consistent with deeper encoding potentially driving the retention advantage~\cite{Craik1975DepthMemory}, although engagement depth can not be teased apart from exposure time in this experimental design.

Exposure time and engagement depth may both be impacted by the AI-Assistance and are inseparable in this experimental procedure. Time on Task was not controlled as participants were given a single 30-minute budget and self-paced through the tasks. The AI-Assisted group finished sooner \emph{because} the tool produced working solutions quickly with little deliberation to the point that some of the AI-Assisted participant's responses could not be interpreted physiologically. For that reason we did not adjust the retention or effort comparisons for time on task. Doing so would condition on a post-treatment variable and would answer a different question from those posed by \textbf{RQ1b} and \textbf{RQ1c}. Equally, the data cannot determine whether the No AI group retained more \emph{because} they spent longer with the material. We therefore present time on task as a descriptive marker of the offloading pattern, consistent with the mental effort and retention results, and leave its causal separation from engagement depth to future work (\autoref{sec:limitations}).

Physiological signals did not identify any significant group level differences between both groups, potentially impacted by data noise discussed later in \autoref{sec:limitations}. In particular, the HRV analysis retained only a minority of task intervals after quality gates. The cardiac insignificance should be read as inconclusive rather than as evidence of no group difference. While not significant, \autoref{fig:pupil_load} does show Task~3 required an increase in effort for the No AI group compared to Task~2, whereas the AI-Assisted group continued to reduce signals of effort and a clearer pattern could have emerged with additional tasks.

CLT offers an explanation for this pattern. The effort GenAI assistance impacted appears to have been germane rather than extraneous or intrinsic. Developing a solution, and retrieving the knowledge for that development, are the processes through which schema form via germane load~\cite{Sweller2011CognitiveTheory}. GenAI driven interactions especially, where the human acts as \emph{reverse centaur} only supplying input to the GenAI tool and blindly accepting the GenAI output, appears to completely side-step the schema formation process~\cite{doctorow2026reverse}. Instead they appear to be entering an offloading loop~\cite{Liu2026ToolEducation} relinquishing that germane processing to the GenAI assistant while the intrinsic load remains, as evidenced by the divergence in mental effort and difficulty in the AI-Assisted group. The No AI participants expended that effort generating or updating schemas and retained more of the content, while AI-Assisted participants reported a flatter effort trajectory and retained less.

\subsection{Ownership and Authorship with AI}
\textbf{RQ1d}
The offloading does not appear limited to cognitive load. AI-Assisted participants reported substantially lower sense of psychological ownership over the submitted work in addition to self-attributing only $45\%$ ownership over the code submitted compared to the $81\%$ ownership by the No AI group. This relinquishing of ownership is supported by the pattern of disengagement observed in the cognitive measurements; it can be hard to feel ownership over something one was not engaged in creating. Retention and ownership are unlikely to be independent outcomes therefore the two results should be taken as convergent evidence of reduced engagement rather than as two separate effects.

This lack of ownership extends to authorship as well. All of the No AI participants would attribute themselves as the primary author of the code, with half either acknowledging Google or listing it as a secondary author. Meanwhile only $7\%$ of the AI-Assisted participants would attribute themselves as the full primary author and $38\%$ relinquish primary authorship to ChatGPT, reflecting each groups sense of ownership over the code. This aligns with previous work finding participants reluctance to attribute authorship to AI generated content~\cite{Draxler2024TheAuthors}, however, Wasi et al.~\cite{Wasi2024LLMsReasoning} found a disconnect on type of content technical vs. creative. It could be that computing students more closely associate writing code as a creative medium than a technical one like writing an email~\cite{Levin2021CodeDesign}. 

All together, the No AI participants both engaged more deeply and were exposed to the task content for longer resulting in lower initial task performance, but higher immediate and delayed retention and a larger sense of ownership over the code submitted. On the other hand AI-Assisted participants completed the assessment in less time with a higher initial task performance, but significantly lower immediate and follow up retention with substantially lower sense of ownership over the code submitted. Taken together, in this sample the short-term performance gain on a single assessment came with a measurable cost to retention and ownership. 

\subsection{Implications and Mitigation}
We put forward potential design implications and suggested mitigation for GenAI assisted learning systems that aim to balance performance support with durable learning at the institution, educator, and student levels.

\subsubsection{Human--AI Learning Design}
AI assisted learning systems should scaffold cognitive development, not replace it. GenAI assisted learning systems should adapt to promote content retention and engage a useful level of cognitive load. GenAI assisted learning systems should utilise current cognitive load theory information~\cite{Sweller2024CognitiveDifferences} as a foundation for processes in determining individualised cognitive load management of users. 

With a user in a state where they are more cognitively ready to learn, these systems should put heavy emphasis on evidence based methods for promoting learning and content retention in digital learning scenarios~\cite{Mayer2024TheLearning}. These systems should also realise that educational interactions are not one off instances and instead these students are expected to retain the information, so these systems should take extra care to support long term retention, not just direct question answering or immediate retention. These interactions should support the act of learning, not just the end result. 

Systems must support active contribution to preserve ownership. Either by actively requiring user input that is integrated into a solution or by eliciting individual involvement in the creation process while being aware of infringing on the students' autonomy~\cite{Xu2024WhatCo-Creations}.

Taken together these design principles can help a GenAI-assisted learning system to maintain students' ownership and content retention while increasing performance. 

\subsubsection{Mitigation}
Institutions should develop policies and procedures that promote individual growth as a key pillar in their institutional framework. Policies that place students in direct competition with each other, like maximum passing or minimum fail thresholds, will create perverse incentives for students to use GenAI not to out-learn but to outscore other students. GenAI then becomes an overly enticing incentive to outscore other students and creates an economy of haves and have not with greater socioeconomic dynamics where students who can afford access to better models will outperform those who cant. 

Teachers should take from this study that GenAI assistance highlights gaps in our current teaching curricula and methods. Primarily, this research shows that GenAI Assistance allows students to out perform those who do not use it on summative assessments. This is a clear advantage, even with free versions of available models. As institutions continue to adopt policies that permit disclosed use~\cite{Qian2026GoverningUniversities,Parker2025ComparativeCountries,Geng2026MappingAnalysis}, keeping GenAI out of the classroom is not a viable stance. However, this research shows evidence that the retention of information is impacted when GenAI assistance is used, which may produce situations where students perform well when GenAI assistance is available yet perform markedly worse in individualised assessments~\cite{Whittford2026BrownCheat}. Therefore the traditional reliance on summative assessments as a proxy for knowledge acquisition needs to be reassessed. We suggest course structures that promote increased opportunities for students to demonstrate individual knowledge such as hands on or in person interactive activities where GenAI assistance is not available. If GenAI usage is integrated into the course, incorporating a "critique-the-AI"~\cite{Liu2026ToolEducation} phase where students validate, test, and justify modification of GenAI output could help engage with germane load processing and possibly enhance sense of ownership and authorship.

Students should be active participants in their education. The short term gain of GenAI Assistance on task performance is gone immediately when asked to recall that information without GenAI assistance. We recommend students attempt to avoid GenAI Assistance when formulating new knowledge scaffolding. If GenAI assistance is compulsory, try and utilise the GenAI assistant to reinforce knowledge scaffolding by having it ask questions, provide similar examples, walk through the steps of the problem; all without simply getting an answer from the assistant. 

This is further exemplified by AI-Assisted participants only attributing ownership of $45\%$ of the code to themselves compared to $81\%$ ownership attribution by the No AI group. Students should interact with GenAI assistants in a manner in which they can retain a sense of ownership of the output, by providing meaningful input and engagement with the content being generated. 

\section{Limitations and Future Work}
\label{sec:limitations}
There were several limitations of this study. Firstly, the participants all self-selected to participate and are all from a single university first-year cohort. This could introduce self-selection bias and these results may not be generalisable to all students. Future studies should look to replicate this experimental design across different geographic and socioeconomic regions. Specifically gender and other traditionally minority group specific focused replications would be beneficial to a more holistic understanding of the impacts of GenAI across gender and minority groups. 

Secondly, this study does not control for how students use GenAI tools to complete the assessment. During the study a pattern was observed where of the $29$ students in the AI-Assisted group, $15$ generated AI driven responses by copying and pasting the instructions or starter code into ChatGPT and accepting the solution from ChatGPT, while $14$ wrote code themselves and used ChatGPT as a validation or debugging tool only. As this was not a priori factor considered in this study it is possible that the different ways in which GenAI tools are being used could impact learning processes differently. Future studies could conduct controlled experiments specifically regimenting how students are instructed to use GenAI tools to further explore if the manner in which GenAI is used has an impact on student learning processes. 

Thirdly, physiological measurements are highly sensitive to variability in participants physiology and behaviour. Eye tracking is sensitive to glasses/lenses, eye physiology, and head distance from the tracking box~\cite{Ezer2024EYEIMPROVEMENTS}. Several eye tracking measurements were excluded from analysis due to issues in capturing the participant's eye consistently, particularly during the baseline as the subsequent task deltas depended on this measurement. These could be participants who wore glasses, had irregular pupil physiology, or participants who shifted during the task leaning toward or away from the screen during the task, or looking up to the roof or down at the keyboard when thinking or typing. Additionally, heart rate variability (HRV) measurements are sensitive to strap tension, sensor placement position, and individual physiology like arm hair and skin tone~\cite{Fine2021SourcesMonitoring}. Many of the heart rate measurements were excluded possibly due to the sensor tension or fit during the experiment and the continuous fine motor demands of typing code on a bicep worn sensor. future research should amend the experiment protocol in an attempt to mitigate potential losses of physiological data such as including a chin rest or other guidance mechanism to ensure the participant's head is in range and visible to the eye tracker or recalibrate the tracker between tasks, or utilising a heart rate capture method that is less sensitive to fine motor movement such as a chest strap sensor or lead-based sensors.  

Additionally, the coding assessment comprised an active typing-test baseline and three coding challenges in a fixed order with a maximum of 30 minutes with no fixed time limits per question and cued recall 48~hours later. The baseline typing test was deliberately chosen over a rest interval so motor and visual elements would be matched to the coding tasks, therefore pupil~$\Delta$s are interpreted in this work as change relative to the baseline anchor. Future work should look to counterbalance or repeat baselines between tasks to estimate position effects on pupil reactions directly. Additional experiment variations that investigate the impact of question order, time on task, duration, and recall prompt timing would be beneficial. 

Finally, it is possible that participants could be less engaged in this experiment than they would be in a graded activity. Future research could investigate the impact of incentives in these interactions by integrating the experiment structure into a course or otherwise incentivise participants perhaps by informing participants their compensation would be tied to performance. 

\section{Conclusion}
Students are using GenAI in their school work to learn. This research expands on the preliminary growing research indicating that there is a negative cognitive impact of GenAI tool usage. Understanding how these GenAI tools impact our cognitive and learning processes now is paramount. The longer we wait to investigate these impacts, the greater the potential risk of losing learning capabilities in future generations grows. 

This research identified GenAI Assistance impacts student learning process in a particularly unique way by providing a real measurable boost to immediate task performance, however it comes at a substantial cost to retention, mental effort, ownership and authorship of the content. In an academic setting, both the educators using assessment performance as a proxy for knowledge and the students getting high marks on those assessments can be lulled into an illusion of learning taking place that shatters upon attempts to recall that knowledge. We feel this is not a sustainable trade off and is detrimental to students.

The design of educational GenAI assistance systems should look to fully support student cognitive development by managing the students cognitive load; not to completely eliminate it, but to utilise it to promote not only initial content acquisition but also ownership and extended recall to develop durable learning in students. We put forth our suggested mitigation for institutions to enact policy that can help foster environments that ensure students remain active participants in their education, for teachers to rethink how we infer knowledge transfer in courses, and for students to remain engaged and invested in their education. 

This research contributes to the emerging body of work on understanding the interplay of GenAI systems and human cognitive learning processes. This field of study will continue to grow in importance as GenAI quickly spreads and integrates with many legacy systems and introduces new systems seeking global adoption. AI's ubiquitous presence must be addressed with caution. 

Much of human history has been looking back and attempting to course correct as damages accumulate years, or decades later. This is an opportunity to develop our understanding of the impacts of this emerging technology in lockstep with this technology's development.

\begin{acks}
This project was supported by a Google.org GAIR/GARA grant which contributed towards the participant compensation.
\end{acks}

\bibliographystyle{ACM-Reference-Format}
\bibliography{base}

@techreport{Pedro2019ArtificialDevelopment,
    title = {{Artificial intelligence in education : challenges and opportunities for sustainable development}},
    year = {2019},
    institution = {UNESCO},
    journal = {MINISTERIO DE EDUCACION},
    author = {Pedro, Francesc and Subosa, Miguel and Rivas, Axel and Valverde, Paula},
    publisher = {UNESCO},
    url = {https://repositorio.minedu.gob.pe/handle/20.500.12799/6533},
    pages ={},
    number = {},
    volume = {},
}

@article{Nguyen2024UnmaskingUndergraduates,
    title = {{Unmasking academic cheating behavior in the artificial intelligence era: Evidence from Vietnamese undergraduates}},
    year = {2024},
    journal = {Education and Information Technologies 2024 29:12},
    author = {Nguyen, Hung Manh and Goto, Daisaku},
    number = {12},
    month = {2},
    pages = {15999--16025},
    volume = {29},
    publisher = {Springer},
    url = {https://link.springer.com/article/10.1007/s10639-024-12495-4},
    isbn = {0123456789},
    doi = {10.1007/S10639-024-12495-4},
    issn = {1573-7608}
}

@article{Oravec2023ArtificialBard,
    title = {{Artificial Intelligence Implications for Academic Cheating: Expanding the Dimensions of Responsible Human-AI Collaboration with ChatGPT and Bard}},
    year = {2023},
    journal = {Journal of Interactive Learning Research},
    author = {Oravec, Jo Ann},
    number = {2},
    pages = {213--237},
    volume = {34},
    publisher = {Association for the Advancement of Computing in Education},
    doi = {10.70725/304731GMMVHW},
    issn = {1093023X}
}

@incollection{Roscoe2022InclusionEducation,
    title = {{Inclusion and equity as a paradigm shift for artificial intelligence in education}},
    year = {2022},
    booktitle = {Artificial Intelligence in STEM Education: The Paradigmatic Shifts in Research, Education, and Technology},
    author = {Roscoe, Rod D. and Salehi, Shima and Nixon, Nia and Worsley, Marcelo and Piech, Chris and Luckin, Rose},
    edition = {1},
    month = {12},
    pages = {359--373},
    publisher = {CRC Press},
    address = {Boca Raton, FL},
    url = {https://www.taylorfrancis.com/chapters/edit/10.1201/9781003181187-28/inclusion-equity-paradigm-shift-artificial-intelligence-education-rod-roscoe-shima-salehi-nia-nixon-marcelo-worsley-chris-piech-rose-luckin},
    isbn = {9781000814712},
    doi = {10.1201/9781003181187-28/INCLUSION-EQUITY-PARADIGM-SHIFT-ARTIFICIAL-INTELLIGENCE-EDUCATION-ROD-ROSCOE-SHIMA-SALEHI-NIA-NIXON-MARCELO-WORSLEY-CHRIS-PIECH-ROSE-LUCKIN}
}

@inproceedings{Ghosh2024ChatGPTClasses,
    title = {{ChatGPT as a Tool for Equitable Education in Engineering Classes}},
    year = {2024},
    booktitle = {2024 ASEE Annual Conference \& Exposition},
    author = {Ghosh, Sourojit},
    month = {6},
    numpages = {16},
    publisher = {American Society for Engineering Education},
    doi = {10.18260/1-2--48458},
    address = {Portland, oregon},
    issn = {21535965}
}

@article{Holstein2021EquityEducation,
    title = {{Equity and Artificial Intelligence in Education: Will "AIEd" Amplify or Alleviate Inequities in Education?}},
    journal      = {CoRR},
    volume       = {abs/2104.12920},
    year = {2021},
    author = {Holstein, Kenneth and Doroudi, Shayan},
    month = {4},
    url = {https://arxiv.org/pdf/2104.12920},
    pages ={},
    arxivId = {2104.12920}
}

@Inbook{Srinivasa2022HarnessingEducation,
    author={Srinivasa, K. G. and Kurni, Muralidhar and Saritha, Kuppala},
    title="Harnessing the Power of AI to Education",
    bookTitle="Learning, Teaching, and Assessment Methods for Contemporary Learners: Pedagogy for the Digital Generation",
    year="2022",
    publisher="Springer Nature Singapore",
    address="Singapore",
    pages="311--342",
    isbn="978-981-19-6734-4",
    doi="10.1007/978-981-19-6734-4_13",
    url="https://doi.org/10.1007/978-981-19-6734-4_13"
}

@misc{Kosmyna2025YourTask,
      title={Your Brain on ChatGPT: Accumulation of Cognitive Debt when Using an AI Assistant for Essay Writing Task}, 
      author={Nataliya Kosmyna and Eugene Hauptmann and Ye Tong Yuan and Jessica Situ and Xian-Hao Liao and Ashly Vivian Beresnitzky and Iris Braunstein and Pattie Maes},
      year={2025},
      eprint={2506.08872},
      archivePrefix={arXiv},
      primaryClass={cs.AI},
      url={https://arxiv.org/abs/2506.08872}, 
}

@article{Paas1992TrainingApproach,
    title = {{Training Strategies for Attaining Transfer of Problem-Solving Skill in Statistics: A Cognitive-Load Approach}},
    year = {1992},
    journal = {Journal of Educational Psychology},
    author = {Paas, Fred G.W.C.},
    number = {4},
    pages = {429--434},
    volume = {84},
    doi = {10.1037/0022-0663.84.4.429},
    issn = {00220663}
}

@article{Paas2003CognitiveTheory,
    title = {{Cognitive load measurement as a means to advance cognitive load theory}},
    year = {2003},
    journal = {Educational Psychologist},
    author = {Paas, Fred and Tuovinen, Juhani E. and Tabbers, Huib and Van Gerven, Pascal W.M.},
    number = {1},
    pages = {63--71},
    volume = {38},
    publisher = {Lawrence Erlbaum Associates Inc.},
    url = {https://www.tandfonline.com/doi/pdf/10.1207/S15326985EP3801_8},
    doi = {10.1207/S15326985EP3801{\_}8;ISSUE:ISSUE:DOI},
    issn = {00461520}
}

@article{Krejtz2018EyeGaze,
    title = {{Eye tracking cognitive load using pupil diameter and microsaccades with fixed gaze}},
    year = {2018},
    journal = {PLOS ONE},
    author = {Krejtz, Krzysztof and Duchowski, Andrew T. and Niedzielska, Anna and Biele, Cezary and Krejtz, Izabela},
    number = {9},
    month = {9},
    pages = {e0203629},
    volume = {13},
    publisher = {Public Library of Science},
    url = {https://journals.plos.org/plosone/article?id=10.1371/journal.pone.0203629},
    isbn = {1111111111},
    doi = {10.1371/JOURNAL.PONE.0203629},
    issn = {1932-6203},
    pmid = {30216385}
}

@article{Gavas2017EstimationDilation,
    title = {{Estimation of cognitive load based on the pupil size dilation}},
    year = {2017},
    journal = {2017 IEEE International Conference on Systems, Man, and Cybernetics, SMC 2017},
    author = {Gavas, Rahul and Chatterjee, Debatri and Sinha, Aniruddha},
    month = {11},
    pages = {1499--1504},
    volume = {2017-January},
    publisher = {Institute of Electrical and Electronics Engineers Inc.},
    url = {https://ieeexplore.ieee.org/abstract/document/8122826},
    isbn = {9781538616451},
    doi = {10.1109/SMC.2017.8122826}
}

@article{Sibley2011PupilLearning,
    author = {Ciara Sibley and Joseph Coyne and Carryl Baldwin},
    title ={Pupil Dilation as an Index of Learning},
    journal = {Proceedings of the Human Factors and Ergonomics Society Annual Meeting},
    volume = {55},
    number = {1},
    pages = {237-241},
    year = {2011},
    doi = {10.1177/1071181311551049},
    URL = { https://doi.org/10.1177/1071181311551049},
    eprint = { https://doi.org/10.1177/1071181311551049}
}

@article{Schweizer2025Wrist-WornStudy.,
    title = {{Wrist-Worn and Arm-Worn Wearables for Monitoring Heart Rate During Sedentary and Light-to-Vigorous Physical Activities: Device Validation Study.}},
    year = {2025},
    journal = {JMIR cardio},
    author = {Schweizer, Theresa and Gilgen-Ammann, Rahel},
    number = {1},
    month = {3},
    pages = {e67110},
    volume = {9},
    publisher = {JMIR Cardio},
    url = {http://www.ncbi.nlm.nih.gov/pubmed/40116771 http://www.pubmedcentral.nih.gov/articlerender.fcgi?artid=PMC11951816},
    doi = {10.2196/67110},
    issn = {2561-1011},
    pmid = {40116771}
}

@article{He2025WhichCo-Creation,
    title = {{Which Contributions Deserve Credit? Perceptions of Attribution in Human-AI Co-Creation}},
    year = {2025},
    journal = {CHI Conference on Human Factors in Computing Systems (CHI '25), April 26-May 1, 2025, Yokohama, Japan},
    author = {He, Jessica and Houde, Stephanie and Weisz, Justin D.},
    month = {2},
    pages = {30},
    volume = {1},
    publisher = {ACM},
    url = {https://arxiv.org/pdf/2502.18357},
    doi = {10.1145/3706598.3713522},
    arxivId = {2502.18357}
}

@article{VanDyne2004PsychologicalBehavior,
    title = {{Psychological ownership and feelings of possession: Three field studies predicting employee attitudes and organizational citizenship behavior}},
    year = {2004},
    journal = {Journal of Organizational Behavior},
    author = {Van Dyne, Linn and Pierce, Jon L.},
    number = {4},
    month = {6},
    pages = {439--459},
    volume = {25},
    publisher = {John Wiley {\&} Sons, Ltd},
    url = {/doi/pdf/10.1002/job.249 https://onlinelibrary.wiley.com/doi/abs/10.1002/job.249 https://onlinelibrary.wiley.com/doi/10.1002/job.249},
    doi = {10.1002/JOB.249;ISSUE:ISSUE:DOI},
    issn = {08943796}
}

@Inbook{Olckers2017MeasuringReview,
    author="Olckers, Chantal and van Zyl, Llewellyn",
    title="Measuring Psychological Ownership: A Critical Review",
    bookTitle="Theoretical Orientations and Practical Applications of Psychological Ownership",
    year="2017",
    publisher="Springer International Publishing",
    address="Cham",
    pages="61--78",
    isbn="978-3-319-70247-6",
    doi="10.1007/978-3-319-70247-6_4",
    url="https://doi.org/10.1007/978-3-319-70247-6_4"
}

@article{Deng2025DoesStudies,
    title = {{Does ChatGPT enhance student learning? A systematic review and meta-analysis of experimental studies}},
    year = {2025},
    journal = {Computers {\&} Education},
    author = {Deng, Ruiqi and Jiang, Maoli and Yu, Xinlu and Lu, Yuyan and Liu, Shasha},
    month = {4},
    pages = {105224},
    volume = {227},
    publisher = {Pergamon},
    url = {https://www.sciencedirect.com/science/article/pii/S0360131524002380},
    doi = {10.1016/J.COMPEDU.2024.105224},
    issn = {0360-1315}
}

@inproceedings{Finnie-Ansley2023MyExercises,
    author = {Finnie-Ansley, James and Denny, Paul and Luxton-Reilly, Andrew and Santos, Eddie Antonio and Prather, James and Becker, Brett A.},
    title = {My AI Wants to Know if This Will Be on the Exam: Testing OpenAI’s Codex on CS2 Programming Exercises},
    year = {2023},
    isbn = {9781450399418},
    publisher = {Association for Computing Machinery},
    address = {New York, NY, USA},
    url = {https://doi.org/10.1145/3576123.3576134},
    doi = {10.1145/3576123.3576134},
    booktitle = {Proceedings of the 25th Australasian Computing Education Conference},
    pages = {97–104},
    numpages = {8},
    location = {Melbourne, VIC, Australia},
    series = {ACE '23}
}

@article{Wasi2024LLMsReasoning,
    title = {{LLMs as Writing Assistants: Exploring Perspectives on Sense of Ownership and Reasoning}},
    year = {2024},
    journal = {ACM International Conference Proceeding Series},
    author = {Wasi, Azmine Toushik and Islam, Mst Rafia and Islam, Raima},
    month = {10},
    pages = {38--42},
    volume = {1},
    publisher = {Association for Computing Machinery},
    url = {https://arxiv.org/pdf/2404.00027v5 http://dx.doi.org/10.1145/3690712.3690723},
    isbn = {9798400710315},
    doi = {10.1145/3690712.3690723}
}

@article{Draxler2024TheAuthors,
    title = {{The AI Ghostwriter Effect: When Users do not Perceive Ownership of AI-Generated Text but Self-Declare as Authors}},
    year = {2024},
    journal = {ACM Transactions on Computer-Human Interaction},
    author = {Draxler, Fiona and Werner, Anna and Lehmann, Florian and Hoppe, Matthias and Schmidt, Albrecht and Buschek, Daniel and Welsch, Robin},
    number = {2},
    month = {2},
    volume = {31},
    publisher = {Association for Computing Machinery},
    url = {https://dl.acm.org/doi/pdf/10.1145/3637875},
    doi = {10.1145/3637875;WGROUP:STRING:ACM},
    issn = {15577325},
    arxivId = {2303.03283},
    pages ={}
}

@article{Paas1994InstructionalTasks,
    title = {{Instructional control of cognitive load in the training of complex cognitive tasks}},
    year = {1994},
    journal = {Educational Psychology Review},
    author = {Paas, Fred G.W.C. and Van Merri{\"{e}}nboer, Jeroen J.G.},
    number = {4},
    month = {12},
    pages = {351--371},
    volume = {6},
    publisher = {Kluwer Academic Publishers-Plenum Publishers},
    url = {https://link.springer.com/article/10.1007/BF02213420},
    doi = {10.1007/BF02213420/METRICS},
    issn = {1040726X}
}

@inproceedings{Lee2025TheWorkers,
    author = {Lee, Hao-Ping (Hank) and Sarkar, Advait and Tankelevitch, Lev and Drosos, Ian and Rintel, Sean and Banks, Richard and Wilson, Nicholas},
    title = {The Impact of Generative AI on Critical Thinking: Self-Reported Reductions in Cognitive Effort and Confidence Effects From a Survey of Knowledge Workers},
    year = {2025},
    isbn = {9798400713941},
    publisher = {Association for Computing Machinery},
    address = {New York, NY, USA},
    url = {https://doi.org/10.1145/3706598.3713778},
    doi = {10.1145/3706598.3713778},
    booktitle = {Proceedings of the 2025 CHI Conference on Human Factors in Computing Systems},
    articleno = {1121},
    numpages = {22},
    location = {    },
    series = {CHI '25}
}

@article{Shaffer2017AnNorms,
    title = {{An Overview of Heart Rate Variability Metrics and Norms}},
    year = {2017},
    journal = {Frontiers in Public Health},
    author = {Shaffer, Fred and Ginsberg, J. P.},
    month = {9},
    pages = {290215},
    volume = {5},
    publisher = {Frontiers Media S.A.},
    url = {www.frontiersin.org},
    doi = {10.3389/FPUBH.2017.00258/FULL},
    issn = {22962565},
    pmid = {29034226}
}

@article{vanderWel2018PupilReview,
    title = {{Pupil dilation as an index of effort in cognitive control tasks: A review}},
    year = {2018},
    journal = {Psychonomic Bulletin {\&} Review 2018 25:6},
    author = {van der Wel, Pauline and van Steenbergen, Henk},
    number = {6},
    month = {2},
    pages = {2005--2015},
    volume = {25},
    publisher = {Springer},
    url = {https://link.springer.com/article/10.3758/s13423-018-1432-y},
    isbn = {25:20052015},
    doi = {10.3758/S13423-018-1432-Y},
    issn = {1531-5320},
    pmid = {29435963}
}

@article{Simmons2011False-positiveSignificant,
    title = {{False-positive psychology: Undisclosed flexibility in data collection and analysis allows presenting anything as significant}},
    year = {2011},
    journal = {Psychological Science},
    author = {Simmons, Joseph P. and Nelson, Leif D. and Simonsohn, Uri},
    number = {11},
    pages = {1359--1366},
    volume = {22},
    publisher = {SAGE Publications Inc.},
    url = {https://journals.sagepub.com/doi/10.1177/0956797611417632},
    doi = {10.1177/0956797611417632;CTYPE:STRING:JOURNAL},
    issn = {14679280},
    pmid = {22006061}
}

@article{Delacre2017WhyT-Test,
    title = {{Why psychologists should by default use welch's t-Test instead of student's t-Test}},
    year = {2017},
    journal = {International Review of Social Psychology},
    author = {Delacre, Marie and Lakens, Daniël and Leys, Christophe},
    number = {1},
    pages = {92--101},
    volume = {30},
    publisher = {Ubiquity Press Ltd},
    doi = {10.5334/IRSP.82},
    issn = {23978570}
}

@article{Lakens2013CalculatingANOVAs,
    title = {{Calculating and reporting effect sizes to facilitate cumulative science: A practical primer for t-tests and ANOVAs}},
    year = {2013},
    journal = {Frontiers in Psychology},
    author = {Lakens, Daniël},
    number = {NOV},
    month = {11},
    pages = {62627},
    volume = {4},
    publisher = {Frontiers},
    url = {www.frontiersin.org},
    doi = {10.3389/FPSYG.2013.00863/TEXT},
    issn = {16641078},
    pmid = {24324449}
}

@unpublished{Krippendorff2011ComputingAlpha-Reliability,
    title = {{Computing Krippendorff's Alpha-Reliability}},
    year = {2011},
    author = {Krippendorff, Klaus},
    number = {43},
    month = {1},
    url = {https://repository.upenn.edu/handle/20.500.14332/2089}
}

@article{krippendorff2004MeasuringData,
    title = {{Measuring the Reliability of Qualitative Text Analysis Data}},
    year = {2004},
    journal = {Quality and Quantity 2004 38:6},
    author = {krippendorff, Klaus},
    number = {6},
    month = {12},
    pages = {787--800},
    volume = {38},
    publisher = {Springer},
    url = {https://link.springer.com/article/10.1007/s11135-004-8107-7},
    doi = {10.1007/S11135-004-8107-7},
    issn = {1573-7845}
}

@article{Holm1979AProcedure,
    title = {{A Simple Sequentially Rejective Multiple Test Procedure}},
    year = {1979},
    journal = {Scandinavian Journal of Statistics},
    author = {Holm, Sture},
    number = {2},
    pages = {65--70},
    volume = {6},
    publisher = {[Board of the Foundation of the Scandinavian Journal of Statistics, Wiley]},
    url = {http://www.jstor.org/stable/4615733},
    issn = {03036898, 14679469}
}

@article{Beatty1982Task-evokedResources,
    title = {{Task-evoked pupillary responses, processing load, and the structure of processing resources}},
    year = {1982},
    journal = {Psychological Bulletin},
    author = {Beatty, Jackson},
    number = {2},
    month = {3},
    pages = {276--292},
    volume = {91},
    doi = {10.1037/0033-2909.91.2.276},
    issn = {00332909},
    pmid = {7071262}
}

@article{Mathot2018SafeData,
    title = {{Safe and sensible preprocessing and baseline correction of pupil-size data}},
    year = {2018},
    journal = {Behavior Research Methods 2018 50:1},
    author = {Math{\^{o}}t, Sebastiaan and Fabius, Jasper and Van Heusden, Elle and Van der Stigchel, Stefan},
    number = {1},
    month = {1},
    pages = {94--106},
    volume = {50},
    publisher = {Springer},
    url = {https://link.springer.com/article/10.3758/s13428-017-1007-2},
    doi = {10.3758/S13428-017-1007-2},
    issn = {1554-3528},
    pmid = {29330763}
}

@article{Kret2018PreprocessingCode,
    title = {{Preprocessing pupil size data: Guidelines and code}},
    year = {2018},
    journal = {Behavior Research Methods 2018 51:3},
    author = {Kret, Mariska E. and Sjak-Shie, Elio E.},
    number = {3},
    month = {7},
    pages = {1336--1342},
    volume = {51},
    publisher = {Springer},
    url = {https://link.springer.com/article/10.3758/s13428-018-1075-y},
    doi = {10.3758/S13428-018-1075-Y},
    issn = {1554-3528},
    pmid = {29992408}
}

@article{Lipponen2019AClassification,
    title = {{A robust algorithm for heart rate variability time series artefact correction using novel beat classification}},
    year = {2019},
    journal = {Journal of Medical Engineering {\&} Technology},
    author = {Lipponen, Jukka A. and Tarvainen, Mika P.},
    number = {3},
    month = {4},
    pages = {173--181},
    volume = {43},
    publisher = {Taylor {\&} Francis},
    url = {https://www.tandfonline.com/doi/pdf/10.1080/03091902.2019.1640306},
    doi = {10.1080/03091902.2019.1640306},
    issn = {1464522X},
    pmid = {31314618}
}

@article{Makowski2021NeuroKit2:Processing,
    title = {{NeuroKit2: A Python toolbox for neurophysiological signal processing}},
    year = {2021},
    journal = {Behavior Research Methods 2021 53:4},
    author = {Makowski, Dominique and Pham, Tam and Lau, Zen J. and Brammer, Jan C. and Lespinasse, François and Pham, Hung and Sch{\"{o}}lzel, Christopher and Chen, S. H.Annabel},
    number = {4},
    month = {2},
    pages = {1689--1696},
    volume = {53},
    publisher = {Springer},
    url = {https://link.springer.com/article/10.3758/s13428-020-01516-y},
    doi = {10.3758/S13428-020-01516-Y},
    issn = {1554-3528},
    pmid = {33528817}
}

@article{Hjortskov2004TheWork,
    title = {{The effect of mental stress on heart rate variability and blood pressure during computer work}},
    year = {2004},
    journal = {European Journal of Applied Physiology 2004 92:1},
    author = {Hjortskov, Nis and Riss{\'{e}}n, Dag and Blangsted, Anne Katrine and Fallentin, Nils and Lundberg, Ulf and S{\o}gaard, Karen},
    number = {1},
    month = {2},
    pages = {84--89},
    volume = {92},
    publisher = {Springer},
    url = {https://link.springer.com/article/10.1007/s00421-004-1055-z},
    doi = {10.1007/S00421-004-1055-Z},
    issn = {1439-6327},
    pmid = {14991326}
}

@article{Tavakol2011MakingAlpha,
    title = {{Making sense of Cronbach's alpha}},
    year = {2011},
    journal = {International Journal of Medical Education},
    author = {Tavakol, Mohsen and Dennick, Reg},
    month = {6},
    pages = {53},
    volume = {2},
    url = {https://pmc.ncbi.nlm.nih.gov/articles/PMC4205511/},
    doi = {10.5116/IJME.4DFB.8DFD},
    issn = {20426372},
    pmid = {28029643}
}

@article{Tarvainen2014KubiosSoftware,
    title = {{Kubios HRV – Heart rate variability analysis software}},
    year = {2014},
    journal = {Computer Methods and Programs in Biomedicine},
    author = {Tarvainen, Mika P. and Niskanen, Juha Pekka and Lipponen, Jukka A. and Ranta-aho, Perttu O. and Karjalainen, Pasi A.},
    number = {1},
    month = {1},
    pages = {210--220},
    volume = {113},
    publisher = {Elsevier},
    url = {https://doi.org/10.1111/j.1469-8986.1997.tb02140.x},
    doi = {10.1016/j.cmpb.2013.07.024},
    issn = {01692607},
    pmid = {24054542}
}

@article{Malik1996HeartUse,
    title = {{Heart rate variability: Standards of measurement, physiological interpretation, and clinical use}},
    year = {1996},
    journal = {Circulation},
    author = {Malik, Marek},
    month = {3},
    pages = {1043--1065},
    volume = {93}
}

@book{Maxwell2004DESIGNINGEdition,
    title = {{DESIGNING EXPERIMENTS AND ANALYZING DATA A MODEL COMPARISON PERSPECTIVE Second Edition}},
    year = {2004},
    author = {Maxwell, Scott E. and Delaney, Harold D.},
    edition = {Second},
    publisher = {Taylor {\&} Fancis},
    address = {New York},
    isbn = {13:978-0-8058-3718-6}
}

@misc{UniversityResources,
    author = {{Department of Industry, Science and Resources}},
    title = {{University enrolment and completion in STEM and other fields | | Department of Industry Science and Resources}},
    url = {https://www.industry.gov.au/publications/stem-equity-monitor/higher-education-data/university-enrolment-and-completion-stem-and-other-fields},
    year = {2026}
}

@article{Satterthwaite1946AnComponents,
    title = {{An Approximate Distribution of Estimates of Variance Components}},
    year = {1946},
    journal = {Biometrics Bulletin},
    author = {Satterthwaite, F. E.},
    number = {6},
    month = {12},
    pages = {110},
    volume = {2},
    publisher = {JSTOR},
    doi = {10.2307/3002019},
    issn = {00994987},
    pmid = {20287815}
}

@article{Kuznetsova2017LmerTestModels,
    title = {{lmerTest Package: Tests in Linear Mixed Effects Models}},
    year = {2017},
    journal = {Journal of Statistical Software},
    author = {Kuznetsova, Alexandra and Brockhoff, Per B. and Christensen, Rune H.B.},
    number = {13},
    month = {12},
    pages = {1--26},
    volume = {82},
    publisher = {American Statistical Association},
    url = {https://www.jstatsoft.org/index.php/jss/article/view/v082i13},
    doi = {10.18637/JSS.V082.I13},
    issn = {1548-7660}
}

@article{Nakagawa2013AModels,
    title = {{A general and simple method for obtaining R2 from generalized linear mixed-effects models}},
    year = {2013},
    journal = {Methods in Ecology and Evolution},
    author = {Nakagawa, Shinichi and Schielzeth, Holger},
    number = {2},
    month = {2},
    pages = {133--142},
    volume = {4},
    publisher = {John Wiley {\&} Sons, Ltd},
    url = {/doi/pdf/10.1111/j.2041-210x.2012.00261.x https://onlinelibrary.wiley.com/doi/abs/10.1111/j.2041-210x.2012.00261.x https://besjournals.onlinelibrary.wiley.com/doi/10.1111/j.2041-210x.2012.00261.x},
    doi = {10.1111/J.2041-210X.2012.00261.X},
    issn = {2041210X}
}

@article{Naznin2025ChatGPTReview,
    title = {{ChatGPT Integration in Higher Education for Personalized Learning, Academic Writing, and Coding Tasks: A Systematic Review}},
    year = {2025},
    journal = {Computers},
    author = {Naznin, Kaberi and Al Mahmud, Abdullah and Nguyen, Minh Thu and Chua, Caslon},
    number = {2},
    month = {2},
    pages = {53},
    volume = {14},
    publisher = {Multidisciplinary Digital Publishing Institute (MDPI)},
    url = {https://www.mdpi.com/2073-431X/14/2/53/htm https://www.mdpi.com/2073-431X/14/2/53},
    doi = {10.3390/COMPUTERS14020053/S1},
    issn = {2073431X}
}

@article{Qian2026GoverningUniversities,
    title = {{Governing generative AI in higher education: emerging policy approaches and support ecosystems at innovative U.S. Universities}},
    year = {2026},
    journal = {International Journal for Educational Integrity 2026 22:1},
    author = {Qian, Yufeng},
    number = {1},
    month = {8},
    pages = {25-},
    volume = {22},
    publisher = {BioMed Central},
    url = {https://link.springer.com/article/10.1007/s40979-026-00233-x},
    doi = {10.1007/S40979-026-00233-X},
    issn = {1833-2595}
}

@article{Parker2025ComparativeCountries,
    title = {{Comparative analysis of artificial intelligence policies in universities across five countries}},
    year = {2025},
    journal = {Discover Computing 2025 28:1},
    author = {Parker, Luke and Loper, A. Jane and Hayes, Josh and Karakas, Alice and White, Steven and Hallman, Heidi},
    number = {1},
    month = {11},
    pages = {267-},
    volume = {28},
    publisher = {Springer},
    url = {https://link.springer.com/article/10.1007/s10791-025-09745-5},
    doi = {10.1007/S10791-025-09745-5},
    issn = {2948-2992}
}

@article{Wu2025Human-generativeMotivation,
    title = {{Human-generative AI collaboration enhances task performance but undermines human’s intrinsic motivation}},
    year = {2025},
    journal = {Scientific Reports 2025 15:1},
    author = {Wu, Suqing and Liu, Yukun and Ruan, Mengqi and Chen, Siyu and Xie, Xiao Yun},
    number = {1},
    month = {4},
    pages = {15105-},
    volume = {15},
    publisher = {Nature Publishing Group},
    url = {https://www.nature.com/articles/s41598-025-98385-2},
    doi = {10.1038/s41598-025-98385-2},
    issn = {2045-2322},
    pmid = {40301425}
}

@article{Noy2023ExperimentalIntelligence,
    title = {{Experimental evidence on the productivity effects of generative artificial intelligence}},
    year = {2023},
    journal = {Science},
    author = {Noy, Shakked and Zhang, Whitney},
    number = {6654},
    month = {7},
    pages = {187--192},
    volume = {381},
    publisher = {American Association for the Advancement of Science},
    url = {/doi/pdf/10.1126/science.adh2586?download=true},
    doi = {10.1126/SCIENCE.ADH2586},
    issn = {10959203},
    pmid = {37440646}
}

@article{Bommasani2021OnModels,
author = {Wiggins, Walter F. and Tejani, Ali S.},
title = {On the Opportunities and Risks of Foundation Models for Natural Language Processing in Radiology},
journal = {Radiology: Artificial Intelligence},
volume = {4},
number = {4},
pages = {e220119},
year = {2022},
doi = {10.1148/ryai.220119},
URL = { https://doi.org/10.1148/ryai.220119},
eprint = { https://doi.org/10.1148/ryai.220119}
}

@article{Mustafa2024AAgenda,
    title = {{A systematic review of literature reviews on artificial intelligence in education (AIED): a roadmap to a future research agenda}},
    year = {2024},
    journal = {Smart Learning Environments 2024 11:1},
    author = {Mustafa, Muhammad Yasir and Tlili, Ahmed and Lampropoulos, Georgios and Huang, Ronghuai and Jandri{\'{c}}, Petar and Zhao, Jialu and Salha, Soheil and Xu, Lin and Panda, Santosh and {Kinshuk} and L{\'{o}}pez-Pernas, Sonsoles and Saqr, Mohammed},
    number = {1},
    month = {12},
    pages = {59-},
    volume = {11},
    publisher = {SpringerOpen},
    url = {https://link.springer.com/article/10.1186/s40561-024-00350-5},
    doi = {10.1186/S40561-024-00350-5},
    issn = {2196-7091}
}

@article{Liang2022HolisticModels,
    title = {{Holistic Evaluation of Language Models}},
    year = {2022},
    journal = {Annals of the New York Academy of Sciences},
    author = {Liang, Percy and Bommasani, Rishi and Lee, Tony and Tsipras, Dimitris and Soylu, Dilara and Yasunaga, Michihiro and Zhang, Yian and Narayanan, Deepak and Wu, Yuhuai and Kumar, Ananya and Newman, Benjamin and Yuan, Binhang and Yan, Bobby and Zhang, Ce and Cosgrove, Christian and Manning, Christopher D. and R{\'{e}}, Christopher and Acosta-Navas, Diana and Hudson, Drew A. and Zelikman, Eric and Durmus, Esin and Ladhak, Faisal and Rong, Frieda and Ren, Hongyu and Yao, Huaxiu and Wang, Jue and Santhanam, Keshav and Orr, Laurel and Zheng, Lucia and Yuksekgonul, Mert and Suzgun, Mirac and Kim, Nathan and Guha, Neel and Chatterji, Niladri and Khattab, Omar and Henderson, Peter and Huang, Qian and Chi, Ryan and Xie, Sang Michael and Santurkar, Shibani and Ganguli, Surya and Hashimoto, Tatsunori and Icard, Thomas and Zhang, Tianyi and Chaudhary, Vishrav and Wang, William and Li, Xuechen and Mai, Yifan and Zhang, Yuhui and Koreeda, Yuta},
    number = {1},
    month = {11},
    pages = {140--146},
    volume = {1525},
    publisher = {John Wiley and Sons Inc},
    url = {https://arxiv.org/pdf/2211.09110},
    doi = {10.1111/nyas.15007},
    issn = {17496632},
    pmid = {37230490},
    arxivId = {2211.09110}
}

@article{Wu2025ADirections,
    title = {{A Survey on LLM-Generated Text Detection: Necessity, Methods, and Future Directions}},
    year = {2025},
    journal = {Computational Linguistics},
    author={Wu, Junchao and Yang, Shu and Zhan, Runzhe and Yuan, Yulin and Chao, Lidia Sam and Wong, Derek Fai},
    number = {1},
    month = {3},
    pages = {275--338},
    volume = {51},
    publisher = {MIT Press},
    url = {https://aclanthology.org/2025.cl-1.8/},
    doi = {10.1162/COLI{\_}A{\_}00549},
    issn = {0891-2017},
    arxivId = {2310.14724}
}

@article{Kazemitabaar2023StudyingProgramming,
    title = {{Studying the effect of AI Code Generators on Supporting Novice Learners in Introductory Programming}},
    year = {2023},
    journal = {Conference on Human Factors in Computing Systems - Proceedings},
    author = {Kazemitabaar, Majeed and Chow, Justin and Ma, Carl Ka To and Ericson, Barbara J. and Weintrop, David and Grossman, Tovi},
    number = {23},
    month = {4},
    pages = {23},
    volume = {1},
    publisher = {Association for Computing Machinery},
    url = {https://dl.acm.org/doi/pdf/10.1145/3544548.3580919},
    isbn = {9781450394215},
    doi = {10.1145/3544548.3580919},
    arxivId = {2302.07427}
}

@article{Wang2025ImpactCourses,
    title = {{Impact of AI-agent-supported collaborative learning on the learning outcomes of University programming courses}},
    year = {2025},
    journal = {Education and Information Technologies 2025 30:12},
    author = {Wang, Haoming and Wang, Chengliang and Chen, Zhan and Liu, Fa and Bao, Chunjia and Xu, Xianlong},
    number = {12},
    month = {3},
    pages = {17717--17749},
    volume = {30},
    publisher = {Springer},
    url = {https://link.springer.com/article/10.1007/s10639-025-13487-8},
    isbn = {0123456789},
    doi = {10.1007/S10639-025-13487-8},
    issn = {1573-7608}
}

@article{Li2025GenerativeProcesses,
    title = {{Generative artificial intelligence-supported programming education: Effects on learning performance, self-efficacy and processes}},
    year = {2025},
    journal = {Australasian Journal of Educational Technology},
    author = {Li, Siran and Liu, Jiangyue and Dong, Qianyan},
    number = {3},
    month = {5},
    pages = {1--25},
    volume = {41},
    publisher = {Australasian Society for Computers in Learning in Tertiary Education},
    url = {https://ajet.org.au/index.php/AJET/article/view/9932},
    doi = {10.14742/AJET.9932},
    issn = {1449-5554}
}

@article{Prather2024TheProgrammers,
    title = {{The Widening Gap: The Benefits and Harms of Generative AI for Novice Programmers}},
    year = {2024},
    journal = {ICER 2024 - ACM Conference on International Computing Education Research},
    author = {Prather, James and Reeves, Brent N. and Leinonen, Juho and Macneil, Stephen and Randrianasolo, Arisoa S. and Becker, Brett A. and Kimmel, Bailey and Wright, Jared and Briggs, Ben},
    month = {8},
    pages = {469--486},
    volume = {1},
    publisher = {Association for Computing Machinery, Inc},
    url = {https://dl.acm.org/doi/pdf/10.1145/3632620.3671116},
    isbn = {9798400704765},
    doi = {10.1145/3632620.3671116;CSUBTYPE:STRING:CONFERENCE},
    arxivId = {2405.17739}
}

@article{Denny2024ComputingAI,
    title = {{Computing Education in the Era of Generative AI}},
    year = {2024},
    journal = {Communications of the ACM},
    author = {Denny, Paul and Prather, James and Becker, Brett A. and Finnie-Ansley, James and Hellas, Arto and Leinonen, Juho and Luxton-Reilly, Andrew and Reeves, Brent N. and Santos, Eddie Antonio and Sarsa, Sami},
    number = {2},
    month = {1},
    pages = {56--67},
    volume = {67},
    publisher = {Association for Computing Machinery},
    url = {https://dl.acm.org/doi/pdf/10.1145/3624720},
    doi = {10.1145/3624720},
    issn = {15577317},
    arxivId = {2306.02608}
}

@misc{Geng2026MappingAnalysis,
      title={Mapping the Emerging Curriculum for AI-Assisted Software Engineering via Syllabus Analysis}, 
      author={Francis Geng and Anshul Shah and Mia Chen and Paul Denny and Juho Leinonen and Bill Griswold and Gerald Soosai Raj and Leo Porter},
      year={2026},
      eprint={2608.05898},
      archivePrefix={arXiv},
      primaryClass={cs.SE},
      url={https://arxiv.org/abs/2608.05898}, 
}

@article{Stadler2024CognitiveInquiry,
    title = {{Cognitive ease at a cost: LLMs reduce mental effort but compromise depth in student scientific inquiry}},
    year = {2024},
    journal = {Computers in Human Behavior},
    author = {Stadler, Matthias and Bannert, Maria and Sailer, Michael},
    month = {11},
    pages = {108386},
    volume = {160},
    publisher = {Pergamon},
    url = {https://www.sciencedirect.com/science/article/pii/S0747563224002541},
    doi = {10.1016/J.CHB.2024.108386},
    issn = {0747-5632}
}

@misc{Whittford2026BrownCheat,
    title = {{Brown Professor Suspects Most of His Class Used AI to Cheat}},
    year = {2026},
    booktitle = {Inside Higher Ed},
    author = {Whittford, Emma},
    month = {7},
    url = {https://www.insidehighered.com/news/faculty/learning-assessment/2026/07/08/brown-professor-suspects-most-his-class-used-ai-cheat}
}

@inproceedings{Xu2024WhatCo-Creations,
    author = {Xu, Yuxin and Cheng, Mengqiu and Kuzminykh, Anastasia},
    title = {What Makes It Mine? Exploring Psychological Ownership over Human-AI Co-Creations},
    year = {2024},
    isbn = {9798400718281},
    publisher = {Association for Computing Machinery},
    address = {New York, NY, USA},
    url = {https://doi.org/10.1145/3670947.3670974},
    doi = {10.1145/3670947.3670974},
    booktitle = {Proceedings of the 50th Graphics Interface Conference},
    articleno = {35},
    numpages = {8},
    location = {Halifax, NS, Canada},
    series = {GI '24}
}

@article{Sweller2024CognitiveDifferences,
    title = {{Cognitive load theory and individual differences}},
    year = {2024},
    journal = {Learning and Individual Differences},
    author = {Sweller, John},
    month = {2},
    pages = {102423},
    volume = {110},
    publisher = {JAI},
    url = {https://www.sciencedirect.com/science/article/pii/S1041608024000165},
    doi = {10.1016/J.LINDIF.2024.102423},
    issn = {1041-6080}
}

@article{Mayer2024TheLearning,
    title = {{The Past, Present, and Future of the Cognitive Theory of Multimedia Learning}},
    year = {2024},
    journal = {Educational Psychology Review 2024 36:1},
    author = {Mayer, Richard E.},
    number = {1},
    month = {1},
    pages = {8-},
    volume = {36},
    publisher = {Springer},
    url = {https://link.springer.com/article/10.1007/s10648-023-09842-1},
    isbn = {0123456789},
    doi = {10.1007/S10648-023-09842-1},
    issn = {1573-336X}
}

@article{Craik1975DepthMemory,
    title = {{Depth of processing and the retention of words in episodic memory}},
    year = {1975},
    journal = {Journal of Experimental Psychology: General},
    author = {Craik, Fergus I. and Tulving, Endel},
    number = {3},
    month = {9},
    pages = {268--294},
    volume = {104},
    doi = {10.1037/0096-3445.104.3.268},
    issn = {00963445}
}

@article{Ezer2024EYEIMPROVEMENTS,
    title = {{EYE TRACKING AS TECHNOLOGY IN EDUCATION: FURTHER INVESTIGATION OF DATA QUALITY AND IMPROVEMENTS}},
    year = {2024},
    journal = {INTED2024 Proceedings},
    author = {Ezer, Timur and Grabinger, Lisa and Hauser, Florian and Staufer, Susanne and Mottok, Jürgen},
    month = {3},
    pages = {2955--2961},
    volume = {1},
    publisher = {IATED},
    isbn = {978-84-09-59215-9},
    doi = {10.21125/inted.2024.0802},
    issn = {2340-1079}
}

@article{Fine2021SourcesMonitoring,
    title = {{Sources of Inaccuracy in Photoplethysmography for Continuous Cardiovascular Monitoring}},
    year = {2021},
    journal = {Biosensors 2021, Vol. 11, Page 126},
    author = {Fine, Jesse and Branan, Kimberly L. and Rodriguez, Andres J. and Boonya-Ananta, Tananant and {Ajmal} and Ramella-Roman, Jessica C. and McShane, Michael J. and Cot{\'{e}}, Gerard L.},
    number = {4},
    month = {4},
    pages = {126},
    volume = {11},
    publisher = {Multidisciplinary Digital Publishing Institute},
    url = {https://www.mdpi.com/2079-6374/11/4/126/htm https://www.mdpi.com/2079-6374/11/4/126},
    doi = {10.3390/BIOS11040126},
    issn = {2079-6374},
    pmid = {33923469}
}

@article{Soderstrom2015LearningReview,
    title = {{Learning Versus Performance: An Integrative Review}},
    year = {2015},
    journal = {Perspectives on Psychological Science},
    author = {Soderstrom, Nicholas C. and Bjork, Robert A.},
    number = {2},
    month = {3},
    pages = {176--199},
    volume = {10},
    publisher = {SAGE Publications Inc.},
    url = {https://scholar.google.com/scholar_url?url=https://journals.sagepub.com/doi/pdf/10.1177/1745691615569000&hl=en&sa=T&oi=ucasa&ct=usl&ei=rPqdasq5Fde46rQP-png2AI&scisig=AIVdB-y2Iez0YmaparIsH9VEZA4v},
    doi = {10.1177/1745691615569000},
    issn = {17456924},
    pmid = {25910388}
}

@article{Slamecka1978ThePhenomenon.,
    title = {{The generation effect: Delineation of a phenomenon.}},
    year = {1978},
    journal = {Journal of Experimental Psychology: Human Learning and Memory},
    author = {Slamecka, Norman J. and Graf, Peter},
    number = {6},
    month = {11},
    pages = {592--604},
    volume = {4},
    publisher = {American Psychological Association (APA)},
    doi = {10.1037/0278-7393.4.6.592},
    issn = {0096-1515}
}

@article{Roediger2011TheRetention,
    title = {{The critical role of retrieval practice in long-term retention}},
    year = {2011},
    journal = {Trends in Cognitive Sciences},
    author = {Roediger, Henry L. and Butler, Andrew C.},
    number = {1},
    month = {1},
    pages = {20--27},
    volume = {15},
    publisher = {Elsevier},
    url = {https://www.cell.com/action/showFullText?pii=S1364661310002081 https://www.cell.com/action/showAbstract?pii=S1364661310002081 https://www.cell.com/trends/cognitive-sciences/abstract/S1364-6613(10)00208-1},
    doi = {10.1016/j.tics.2010.09.003},
    issn = {13646613},
    pmid = {20951630}
}

@article{Bjork2011MakingLearning,
    title = {{Making Things Hard on Yourself, But in a Good Way: Creating Desirable Difficulties to Enhance Learning}},
    year = {2011},
    journal = {Psychology and the real world: Essays illustrating fundamental contributions to society},
    author = {Bjork, Elizabeth L and Bjork, Robert and Roediger, Henry L and Mcdermott, Kathleen B and Mcdaniel, Mark A},
    pages = {56--4},
    volume = {2}
}

@article{Bjork2020DesirablePractice,
    title = {{Desirable Difficulties in Theory and Practice}},
    year = {2020},
    journal = {Journal of Applied Research in Memory and Cognition},
    author = {Bjork, Robert A. and Bjork, Elizabeth L.},
    number = {4},
    month = {12},
    pages = {475--479},
    volume = {9},
    publisher = {Elsevier Inc.},
    doi = {10.1016/J.JARMAC.2020.09.003},
    issn = {22113681}
}

@article{Sweller2019CognitiveTechnology,
    title = {{Cognitive load theory and educational technology}},
    year = {2019},
    journal = {Educational Technology Research and Development 2019 68:1},
    author = {Sweller, John},
    number = {1},
    month = {8},
    pages = {1--16},
    volume = {68},
    publisher = {Springer},
    url = {https://link.springer.com/article/10.1007/s11423-019-09701-3},
    isbn = {0123456789},
    doi = {10.1007/S11423-019-09701-3},
    issn = {1556-6501}
}

@article{Sweller1998CognitiveDesign,
    title = {{Cognitive Architecture and Instructional Design}},
    year = {1998},
    journal = {Educational Psychology Review},
    author = {Sweller, John and Van Merrienboer, Jeroen J.G. and Paas, Fred G.W.C.},
    number = {3},
    pages = {251--296},
    volume = {10},
    publisher = {Kluwer Academic/Plenum Publishers},
    url = {https://link.springer.com/article/10.1023/A:1022193728205},
    doi = {10.1023/A:1022193728205/METRICS},
    issn = {1040726X}
}

@article{Sweller2011CognitiveTheory,
    title = {{Cognitive Load Theory}},
    year = {2011},
    journal = {Psychology of Learning and Motivation - Advances in Research and Theory},
    author = {Sweller, John},
    month = {1},
    pages = {37--76},
    volume = {55},
    publisher = {Academic Press},
    url = {https://www.sciencedirect.com/science/chapter/bookseries/pii/B9780123876911000028},
    doi = {10.1016/B978-0-12-387691-1.00002-8},
    issn = {0079-7421}
}

@book{Levin2021CodeDesign,
  title={Code as creative medium: a handbook for computational art and design},
  author={Levin, Golan and Brain, Tega},
  year={2021},
  address={Boston,MA},
  publisher={MIT Press}
}

@article{Liu2026ToolEducation,
    title = {{Tool, tutor, or crutch?: A grounded theory of cognitive scaffolding and offloading in AI-assisted programming education}},
    year = {2026},
    journal = {International Journal of STEM Education 2026 13:1},
    author = {Liu, Dandan and Fan, Guangrui and Pan, Lihu},
    number = {1},
    month = {3},
    pages = {10-},
    volume = {13},
    publisher = {SpringerOpen},
    url = {https://link.springer.com/article/10.1186/s40594-025-00592-w},
    doi = {10.1186/S40594-025-00592-W},
    issn = {2196-7822}
}

@article{Prather2023TheEducation,
    title = {{The Robots are Here: Navigating the Generative AI Revolution in Computing Education}},
    year = {2023},
    journal = {ITiCSE-WGR 2023 - Proceedings of the 2023 Working Group Reports on Innovation and Technology in Computer Science Education},
    author = {Prather, James and Denny, Paul and Leinonen, Juho and Becker, Brett A. and Albluwi, Ibrahim and Craig, Michelle and Keuning, Hieke and Kiesler, Natalie and Kohn, Tobias and Luxton-Reilly, Andrew and MacNeil, Stephen and Petersen, Andrew and Pettit, Raymond and Reeves, Brent N. and Savelka, Jaromir},
    month = {12},
    pages = {108--159},
    volume = {1},
    publisher = {Association for Computing Machinery, Inc},
    url = {https://dl.acm.org/doi/pdf/10.1145/3623762.3633499},
    isbn = {9798400704055},
    doi = {10.1145/3623762.3633499;CSUBTYPE:STRING:CONFERENCE},
    arxivId = {2310.00658}
}

@article{Finnie-Ansley2022TheProgramming,
    title = {{The robots are coming: Exploring the implications of OpenAI codex on introductory programming}},
    year = {2022},
    journal = {ACM International Conference Proceeding Series},
    author = {Finnie-Ansley, James and Denny, Paul and Becker, Brett A. and Luxton-Reilly, Andrew and Prather, James},
    month = {2},
    pages = {10--19},
    volume = {22},
    publisher = {Association for Computing Machinery},
    url = {https://dl.acm.org/doi/pdf/10.1145/3511861.3511863},
    isbn = {9781450396431},
    doi = {10.1145/3511861.3511863}
}

@article{Yilmaz2023TheMotivation,
    title = {{The effect of generative artificial intelligence (AI)-based tool use on students' computational thinking skills, programming self-efficacy and motivation}},
    year = {2023},
    journal = {Computers and Education: Artificial Intelligence},
    author = {Yilmaz, Ramazan and Karaoglan Yilmaz, Fatma Gizem},
    month = {1},
    pages = {100147},
    volume = {4},
    publisher = {Elsevier},
    url = {https://www.sciencedirect.com/science/article/pii/S2666920X23000267},
    doi = {10.1016/J.CAEAI.2023.100147},
    issn = {2666-920X}
}

@article{Alanazi2025TheMeta-Analysis,
    title = {{The Influence of Artificial Intelligence Tools on Learning Outcomes in Computer Programming: A Systematic Review and Meta-Analysis}},
    year = {2025},
    journal = {Computers},
    author = {Alanazi, Manal and Soh, Ben and Samra, Halima and Li, Alice},
    number = {5},
    month = {5},
    pages = {185},
    volume = {14},
    publisher = {Multidisciplinary Digital Publishing Institute (MDPI)},
    url = {https://www.mdpi.com/2073-431X/14/5/185/htm https://www.mdpi.com/2073-431X/14/5/185},
    doi = {10.3390/COMPUTERS14050185/S1},
    issn = {2073431X}
}

@book{doctorow2026reverse,
  title={The reverse centaur's guide to life after AI: how to think about artificial intelligence before it's too late},
  author={Doctorow, Cory},
  year={2026},
  publisher={Verso Books},
  address={207 East 32nd street, New York, NY 10016},
  isbn={9781836745549}
}

@article{rosenbaum1984TheTreatment,
  author  = {Rosenbaum, Paul R.},
  title   = {The Consequences of Adjustment for a Concomitant Variable That Has Been Affected by the Treatment},
  journal = {Journal of the Royal Statistical Society. Series A (General)},
  volume  = {147}, number = {5}, pages = {656--666}, year = {1984}, doi = {10.2307/2981697}
}

@article{montgomery2018HowIt,
  author  = {Montgomery, Jacob M. and Nyhan, Brendan and Torres, Michelle},
  title   = {How Conditioning on Posttreatment Variables Can Ruin Your Experiment and What to Do about It},
  journal = {American Journal of Political Science},
  volume  = {62}, number = {3}, pages = {760--775}, year = {2018}, doi = {10.1111/ajps.12357}
}

@article{miller2001MisunderstandingCovariance,
  author  = {Miller, Gregory A. and Chapman, Jean P.},
  title   = {Misunderstanding Analysis of Covariance},
  journal = {Journal of Abnormal Psychology},
  volume  = {110}, number = {1}, pages = {40--48}, year = {2001}, doi = {10.1037/0021-843X.110.1.40}
}

\appendix
\section{Survey Instrument}
\subsection{Sleep and Caffeine}
In the last 24 hours; did you sleep significantly less than in a regular night?
Yes/No

In the last 24 hours; did you have more caffeine than on a regular day?
Yes/No

\subsection{Experience}
\subsubsection{General Programming Experience}
How much experience do you have programming in any programming language?
\begin{itemize}
    \item 0 Years (No experience)
    \item 1 Year
    \item 2 Years
    \item 3 Years
    \item 4 Years
    \item 5+ Years
\end{itemize}

\subsubsection{C Programming Experience}
How much experience do you have programming in the C programming language?
\begin{itemize}
    \item 0 Years (No experience)
    \item 1 Year
    \item 2 Years
    \item 3 Years
    \item 4 Years
    \item 5+ Years
\end{itemize}

\subsubsection{General LLM Experience}
How much experience do you have using Large Language Models (LLMs)?
\begin{itemize}
    \item None at all
    \item A little
    \item A moderate amount
    \item A lot
    \item A great deal
\end{itemize}

\subsubsection{ChatGPT Experience}
How much experience do you have using ChatGPT?
\begin{itemize}
    \item None at all
    \item A little
    \item A moderate amount
    \item A lot
    \item A great deal
\end{itemize}

\subsection{Authorship Question}
\label{sec:Authorship}
Authorship question adapted from He et al.~\cite{He2025WhichCo-Creation}.

For the next question please refer to the following definitions: 
\begin{itemize}
    \item Sole Developer designed the logic, wrote the code, and fixed errors.
    \item Primary Developer defined the structure and logic and significantly modified code included by others.
    \item Secondary Developer only tweaked parameters, fixed bugs, or included code that was significantly modified by others.
    \item Acknowledged Developer's contribution was helpful like spotting a bug or suggesting a library but not fundamental to the final code. This would be a "Special Thanks".
    \item Co-Developer is a true partnership. Both parties contributed roughly equally to the final functioning code.
\end{itemize}

Imagine you were going to publish this code. Thinking only of the code for the functions submitted; please indicate what you think is the most accurate way to attribute authorship of the code submitted. Remember that there are no right or wrong answers.

Participants selected one of the following options:
\begin{itemize}
    \item You are the sole developer
    \item You are the primary developer; Google/ChatGPT is acknowledged but not listed as a developer
    \item You are the primary developer; Google/ChatGPT is the secondary developer
    \item You and Google/ChatGPT are Co-Developers and have equal contribution
    \item Google/ChatGPT is the primary developer; you are the secondary developer
    \item Google/ChatGPT is the primary developer; you are acknowledged but not as a developer
    \item Google/ChatGPT is the sole developer
\end{itemize}

\subsection{Self-reported Ownership Attribution}
Question modified from Kosmyna et al.~\cite{Kosmyna2025YourTask}.

Thinking only of the code for the functions submitted; how much was yours, how much was (Google/ChatGPT)? Please indicate what percentage you feel was yours.

1--100 point scale

\subsection{Psychological Ownership Scale}
Adapted from Van Dyne and Pierce~\cite{VanDyne2004PsychologicalBehavior}.

Scale:
\begin{itemize}
    \item Strongly Disagree
    \item Somewhat Disagree
    \item Neither agree nor disagree
    \item Somewhat agree
    \item Strongly Agree
\end{itemize}

Thinking only of the code for the functions submitted; Indicate the degree to which you personally agree or disagree with the following statements.
\begin{itemize}
    \item This is MY code
    \item I sense that this is MY code
    \item I feel a very high degree of personal ownership for this code
    \item It is hard for me to think about this code as MINE
\end{itemize}

\end{document}